\documentclass[aps,prd,          groupedaddress,amsmath,amssymb]{revtex4}

\usepackage{epsfig}
\usepackage{dcolumn}
\usepackage{bm}

\begin{document}

\def\d{{\rm d}}
\def\eps{\varepsilon}
\def\mg{m_{\tilde{g}}}
\def\mba{m_{\tilde{b}_1}}
\def\mbb{m_{\tilde{b}_2}}
\newcommand{\fmslash}[1]{\displaystyle{\not}#1}
\def\lp{\left. }
\def\rp{\right. }
\def\lr{\left( }
\def\rr{\right) }
\def\le{\left[ }
\def\re{\right] }
\def\lg{\left\{ }
\def\rg{\right\} }
\def\lb{\left| }
\def\rb{\right| }
\def\beq{\begin{equation}}
\def\eeq{\end{equation}}
\def\bea{\begin{eqnarray}}
\def\eea{\end{eqnarray}}

\preprint{FREIBURG PHENO-07-07}
\preprint{LPSC 07-151}
\title{Gaugino-pair production in polarized and unpolarized hadron collisions}
\author{Jonathan Debove$^a$}
\author{Benjamin Fuks$^b$}
\author{Michael Klasen$^a$}
\email[]{klasen@lpsc.in2p3.fr}
\affiliation{$^a$ Laboratoire de Physique Subatomique et de Cosmologie,
 Universit\'e Joseph Fourier/CNRS-IN2P3/INPG,
 53 Avenue des Martyrs, F-38026 Grenoble, France \\
 $^b$ Physikalisches Institut, Albert-Ludwigs-Universit\"at
 Freiburg, Hermann-Herder-Stra\ss{}e 3, D-79106 Freiburg i.Br.,
 Germany}
\date{\today}
\begin{abstract}
We present an exploratory study of gaugino-pair production in polarized
and unpolarized
hadron collisions, focusing on the correlation of beam polarization and
gaugino/Higgsino mixing in the general Minimal Supersymmetric Standard
Model. Helicity-dependent cross sections induced by neutral and charged
electroweak currents and squark exchanges are computed analytically
in terms of generalized charges, defined similarly for chargino-pair,
neutralino-chargino associated, and neutralino-pair production. Our results
confirm and extend those obtained previously for negligible Yukawa couplings
and nonmixing squarks. Assuming that the lightest chargino mass is known,
we show numerically that measurements of the longitudinal single-spin
asymmetry at the existing polarized $pp$ collider RHIC and at possible
polarization upgrades of the Tevatron or the LHC would allow for a
determination of the gaugino/Higgsino fractions of charginos and
neutralinos. The theoretical uncertainty coming from factorization scale and
squark mass variations and the expected experimental error on the lightest
chargino mass is generally smaller than the one induced by the
polarized parton densities, so that more information on the latter would
considerably improve on the analysis.
\end{abstract}
\maketitle


\vspace*{-100mm}
\noindent FREIBURG PHENO-07-07\\
\noindent LPSC 07-151\\
\vspace*{90mm}

\section{Introduction}
\label{sec:1}

Weak-scale Supersymmetry (SUSY) continues to be a theoretically attractive
extension of the Standard Model (SM) of particle physics. If $R$-parity is
conserved, it provides in particular a convincing candidate for the large
amount of cold dark matter observed in the Universe. In the Minimal
Supersymmetric SM (MSSM) \cite{Haber:1984rc}, this is generally the lightest
neutralino, one of the spin-1/2 SUSY partners of the electroweak gauge
bosons (gauginos) and of the Higgs bosons (Higgsinos), which mix to form
four neutral (neutralino) and two charged (chargino) mass eigenstates. The
gaugino/Higgsino decomposition of the neutralinos/charginos contains
important information about the SUSY-breaking mechanism and plays a crucial
role in the determination of the dark matter relic density
\cite{Baer:2007xd,Herrmann:2007ku}.

Unfortunately, SUSY particles have yet to be found at high-energy
accelerators. The LEP and Tevatron colliders have constrained the gauginos
and scalar partners of the fermions (squarks/sleptons) to be heavier than a
few tens and hundreds of GeV, respectively, and the search for SUSY
particles has thus become one of the defining tasks of the LHC. At the same
time of the LHC start-up, RHIC is scheduled to operate in the years 2009
through 2012 in its polarized $pp$ mode at an increased center-of-mass
energy of $\sqrt{S}=500$ GeV and with a large integrated luminosity of 266
pb$^{-1}$ during each of the ten-week physics runs. It is therefore
interesting to investigate the influence of proton beam polarization on
production cross sections and longitudinal spin asymmetries for SUSY
particle production at the existing polarized $pp$ collider RHIC
\cite{rhicspin} and at possible polarization upgrades of the Tevatron
\cite{Baiod:1995eu} or the LHC \cite{roeck}.

While electron beams in high-energy accelerators can polarize automatically
due to the emission of spin-flip synchrotron radiation via the
Sokolov-Ternov effect, electrons of lower energy and protons have to be
polarized in a source. These beams then have to be accelerated and stored
with little loss of polarization. Today polarized proton beams can be
produced either by a polarized atomic beam source or in an optically
pumped polarized ion source. Pulsed beams with polarization of up to
87\% for a 1 mA $H^-$ beam current and up to 60\% for 5 mA, respectively,
have been achieved with these sources \cite{barber}. Crossing resonances
while accelerating the beam can, however, in principle lead to a reduction of
the polarization, but this can be avoided with a spin rotation using Siberian
snakes.

In a previous paper, we studied the correlation of beam polarization and the
mixing of the scalar partners of left- and right-handed leptons produced in
polarized hadron collisions \cite{Bozzi:2004qq}. Unpolarized production
cross sections for sleptons are known at next-to-leading order (NLO) of
perturbation theory \cite{Beenakker:1999xh} and have recently been resummed
at next-to-leading logarithmic (NLL) accuracy \cite{Bozzi:2006fw,%
Bozzi:2007qr,Bozzi:2007tea}. Here, we present an exploratory study of
gaugino-pair production in polarized hadron collisions, focusing on the
correlation of beam polarization and gaugino/Higgsino mixing in the general
MSSM. Unpolarized cross sections for gaugino pairs are also known at NLO
\cite{Beenakker:1999xh} and have been re-examined recently for possible
signals of non-minimal flavor violation \cite{Bozzi:2007me}. The impact of
gaugino/Higgsino mixing on beam polarization asymmetries in chargino and
neutralino production has been considered previously, albeit only at
$e^+e^-$ colliders, where the asymmetries were found to strongly constrain
the $B$-ino mass parameter $M_1$ or, through the Grand Unified Theory (GUT)
relation $M_1={5\over3} \tan^2\theta_W M_2\simeq0.5 M_2$, equivalently the
$W$-ino mass parameter $M_2$ ($\theta_W$ is the electroweak mixing angle)
\cite{MoortgatPick:2000uz}.

The remainder of this paper is organized as follows: In Sec.\ \ref{sec:2},
we compute analytically the helicity-dependent cross sections induced by
neutral and charged electroweak currents and squark exchanges in terms of
generalized charges. In Sec.\ \ref{sec:3}, we show numerically that
measurements of the longitudinal single-spin asymmetry would allow for a
determination of the gaugino/Higgsino fractions of charginos and
neutralinos, assuming that the lightest chargino mass is known. We also
estimate the theoretical uncertainties coming from variations of the
unphysical factorization scale, the yet unknown squark masses, 
the expected experimental error on the lightest chargino mass, and our
limited knowledge of the polarized parton densities. We summarize our
results in Sec.\ \ref{sec:4}.

\section{Analytical results}
\label{sec:2}

We start by computing analytically the partonic cross sections for the
pair production of gauginos and Higgsinos, whose mixing to neutralinos and
charginos $\tilde{\chi}^{0,\pm}_i$ is described in App.\ \ref{sec:a}, in
terms of generalized electroweak couplings, which are defined in App.\
\ref{sec:b}. The process
\bea
 q(h_a,p_a)\, \bar{q}^\prime(h_b,p_b) \to
 \tilde{\chi}^{0,\pm}_i(p_1)\, \tilde{\chi}^{0,\mp}_j(p_2)
\eea
is induced by initial quarks $q$ and antiquarks $\bar{q}'$ with definite
helicities $h_{a,b}$ and momenta $p_{a,b}$ and is mediated by $s$-channel
electroweak gauge-boson and $t$- and $u$-channel squark exchanges (see
%
\begin{figure}
 \centering
 \epsfig{file=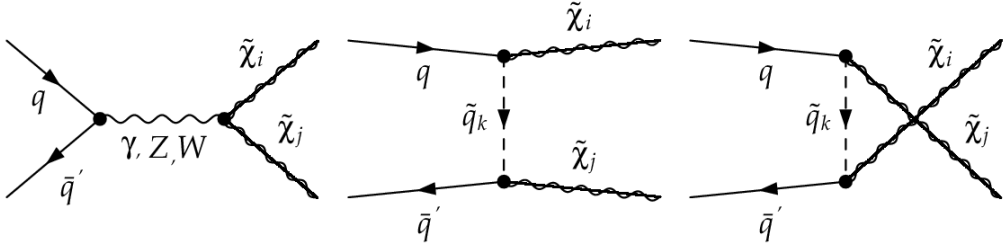,width=.75\columnwidth}
 \caption{\label{fig:1}Tree-level Feynman diagrams for the production
          of gaugino pairs.}
\end{figure}
%
Fig.\ \ref{fig:1}). Its cross section can be expressed generically as
\bea
 \frac{\d \hat{\sigma}^{q\bar{q}'}_{h_a, h_b}}{\d t} &=&
 \frac{\pi \alpha^2}{3 s^2}(1-h_a) (1+h_b) \Big[ \left| Q^u_{LL}
 \right|^2u_{\tilde{\chi}_i} u_{\tilde{\chi}_j} + \left| Q_{LL}^t
 \right|^2t_{\tilde{\chi}_i} t_{\tilde{\chi}_j} + 2~{\rm Re}
 [Q_{LL}^{u\ast} Q_{LL}^t] m_{\tilde{\chi}_{i}}
 m_{\tilde{\chi}_{j}} s \Big] \nonumber \\ & +& \frac{\pi
 \alpha^2}{3 s^2}(1+h_a) (1-h_b) \Big[ \left| Q_{RR}^u \right|^2
 u_{\tilde{\chi}_i} u_{\tilde{\chi}_j} + \left| Q_{RR}^t \right|^2
 t_{\tilde{\chi}_i} t_{\tilde{\chi}_j} + 2~{\rm Re} [Q_{RR}^{u\ast}
 Q_{RR}^t] m_{\tilde{\chi}_{i}} m_{\tilde{\chi}_{j}} s
 \Big]\nonumber \\
 & +& \frac{\pi \alpha^2}{3 s^2}(1-h_a) (1-h_b) \Big[ \left| Q_{LR}^u
 \right|^2 u_{\tilde{\chi}_i} u_{\tilde{\chi}_j} + \left| Q_{LR}^t
 \right|^2 t_{\tilde{\chi}_i} t_{\tilde{\chi}_j} + 2~{\rm Re}
 [Q_{LR}^{u\ast} Q_{LR}^t] (u t - m^2_{\tilde{\chi}_{i}}
 m^2_{\tilde{\chi}_{j}}) \Big]\nonumber\\
 & +& \frac{\pi \alpha^2}{3 s^2}(1+h_a) (1+h_b)
 \Big[ \left| Q_{RL}^u \right|^2 u_{\tilde{\chi}_i}
 u_{\tilde{\chi}_j} + \left| Q_{RL}^t \right|^2 t_{\tilde{\chi}_i}
 t_{\tilde{\chi}_j} + 2~{\rm Re} [Q_{RL}^{u\ast} Q_{RL}^t] (u t -
 m^2_{\tilde{\chi}_{i}} m^2_{\tilde{\chi}_{j}}) \Big],
 \label{eq:xsecgg}
\eea
i.e.\ in terms of generalized charges $Q_{IJ}^{t,u}$, the conventional
Mandelstam variables
\bea
 s=(p_a+p_b)^2,~ t=(p_a-p_1)^2 \mbox{,~and~} u=(p_a-p_2)^2,
\eea
the gaugino and squark masses $m_{\tilde{\chi}^{0,\pm}_{i,j}}$ and
$m_{\tilde{q}_k}$, and the masses of the neutral and charged electroweak
gauge bosons $m_Z$ and $m_W$. Propagators appear as mass-subtracted
Mandelstam variables,
\bea
 \begin{array}{l c l c l c l c}
 s_z &=& s-m_Z^2&,~& s_w &=& s-m_W^2 &,\\[1mm]
 t_{\tilde{q}_k} &=& t-m_{\tilde{q}_k}^2 &,~& u_{\tilde{q}_k} &=&
 u-m_{\tilde{q}_k}^2 &, \\[1mm]
 t_{\tilde{\chi}_i} &=& t-m_{\tilde{\chi}_i}^2 &,~& u_{\tilde{\chi}_i} &=&
 u-m_{\tilde{\chi}_i}^2 &, \end{array}&&
\eea
and the weak interaction is defined through the square of its coupling
constant $g^2=e^2/\sin^2\theta_W$ in terms of the electromagnetic
fine structure constant $\alpha=e^2/(4\pi)$ and the squared sine of the
electroweak mixing angle $x_W=\sin^2\theta_W=s_W^2 = 1-\cos^2\theta_W =
1-c_W^2$.
Unpolarized cross sections, averaged over initial spins, can easily be
derived from the expression
\bea
 \d\hat{\sigma}=\frac{\d\hat{\sigma}_{ 1, 1} + \d\hat{\sigma}_{
 1,-1} + \d\hat{\sigma}_{-1, 1} + \d\hat{\sigma}_{-1,-1}}{4},
 \label{eq:5}
\eea
while single- and double-polarized cross sections, including the same
average factor for initial spins, are given by
\bea
 \d\Delta\hat{\sigma}_L=\frac{\d\hat{\sigma}_{ 1, 1} \pm
 \d\hat{\sigma}_{1,-1} \mp \d\hat{\sigma}_{-1, 1} -
 \d\hat{\sigma}_{-1,-1}}{4} & ~~~{\rm and}~~~ &
 \d\Delta\hat{\sigma}_{LL}=\frac{\d\hat{\sigma}_{ 1, 1} -
 \d\hat{\sigma}_{ 1,-1} - \d\hat{\sigma}_{-1, 1} +
 \d\hat{\sigma}_{-1,-1}}{4},
 \label{eq:6}
\eea
where the upper (lower) signs refer to polarized (anti-)quarks.
The partonic single- and double-spin asymmetries then become
\bea
 \hat{A}_L = \frac{\d\Delta\hat{\sigma}_L}{\d\hat{\sigma}}& {\rm ~~~and~~~}&
 \hat{A}_{LL} = \frac{\d\Delta\hat{\sigma}_{LL}}{\d\hat{\sigma}}.
\eea

For $\tilde{\chi}_i^+\tilde{\chi}_j^-$-production, the generalized charges
are given by
\bea
 Q_{LL}^{u+-} &=& \frac{1}{x_W\,(1-x_W)} \left[
   \frac{L_{Zqq'} O^{\prime L}_{ij}}{s_z}
 + \sum_{k=1}^6 \frac{
 L_{\tilde{\chi}_i^+\tilde{u}_k q^\prime}
 L^\ast_{\tilde{\chi}_j^+\tilde{u}_k q}
 }{u_{\tilde{u}_k}}\right]
 -\frac{e_q \delta_{ij} \delta_{qq^\prime}}{s},~\nonumber\\
 Q_{LL}^{t+-} &=& \frac{1}{x_W\,(1-x_W)} \left[
   \frac{L_{Zqq'} O^{\prime R}_{ij}}{s_z}
 - \sum_{k=1}^6 \frac{
 L^\ast_{\tilde{\chi}_i^+\tilde{d}_k q}
 L_{\tilde{\chi}_j^+\tilde{d}_k q^\prime}
 }{t_{\tilde{d}_k}}\right]
 -\frac{e_q \delta_{ij} \delta_{qq^\prime}}{s},~\nonumber\\
 Q_{RR}^{u+-} &=& \frac{1}{x_W\,(1-x_W)} \left[
   \frac{R_{Zqq'} O^{\prime R}_{ij}}{s_z}
 + \sum_{k=1}^6 \frac{
 R_{\tilde{\chi}_i^+\tilde{u}_k q^\prime}
 R^\ast_{\tilde{\chi}_j^+\tilde{u}_k q}
 }{u_{\tilde{u}_k}}\right]
 -\frac{e_q \delta_{ij} \delta_{qq^\prime}}{s},~\nonumber\\
 Q_{RR}^{t+-} &=& \frac{1}{x_W\,(1-x_W)} \left[
   \frac{R_{Zqq'} O^{\prime L}_{ij}}{s_z}
 - \sum_{k=1}^6 \frac{
 R^\ast_{\tilde{\chi}_i^+\tilde{d}_k q}
 R_{\tilde{\chi}_j^+\tilde{d}_k q^\prime}
 }{t_{\tilde{d}_k}}\right]
 -\frac{e_q \delta_{ij} \delta_{qq^\prime}}{s},~\nonumber\\
 Q_{LR}^{u+-} &=& \frac{1}{x_W\,(1-x_W)}
 \sum_{k=1}^6 \frac{
 R_{\tilde{\chi}_i^+\tilde{u}_k q^\prime}
 L^\ast_{\tilde{\chi}_j^+\tilde{u}_k q}
 }{u_{\tilde{u}_k}},~ \nonumber\\
 Q_{LR}^{t+-} &=& \frac{1}{x_W\,(1-x_W)}
 \sum_{k=1}^6 \frac{
 L^\ast_{\tilde{\chi}_i^+\tilde{d}_k q}
 R_{\tilde{\chi}_j^+\tilde{d}_k q^\prime}
 }{t_{\tilde{d}_k}},~ \nonumber\\
 Q_{RL}^{u+-} &=& \frac{1}{x_W\,(1-x_W)}
 \sum_{k=1}^6 \frac{
 L_{\tilde{\chi}_i^+\tilde{u}_k q^\prime}
 R^\ast_{\tilde{\chi}_j^+\tilde{u}_k q}
 }{u_{\tilde{u}_k}} ,~ \nonumber\\
 Q_{RL}^{t+-} &=& \frac{1}{x_W\,(1-x_W)}
 \sum_{k=1}^6 \frac{
 R^\ast_{\tilde{\chi}_i^+\tilde{d}_k q}
 L_{\tilde{\chi}_j^+\tilde{d}_k q^\prime}
 }{t_{\tilde{d}_k}}.
 \label{eq:8}
\eea
Note that there is no interference between $t$- and $u$-channel diagrams due
to (electromagnetic) charge conservation. After accounting for our
harmonization of generalized charge definitions, which are now similar for
all gaugino channels, our results agree with those published in Ref.\
\cite{Bozzi:2007me} for $\tilde{\chi}_i^-\tilde{\chi}_j^+$-production.
The cross section for chargino-pair production in $e^+e^-$-collisions can be
deduced by setting $e_q \to e_l = -1$, $L_{Zqq'}\to L_{Zee} = T^{3}_l -
e_l\,x_W$ and $R_{Zqq'}\to R_{Zee} = - e_l\,x_W$. Neglecting all Yukawa
couplings, we can then reproduce the calculations of Ref.\
\cite{Choi:1998ei}.

The charges of the neutralino-chargino associated production are given by
\bea
 Q_{LL}^{u0-} &=&
 \frac{1}{x_W\,(1-x_W)} \left[ \frac{L_{Wqq'} O^{L}_{ij}}{s_w}
 + \sum_{k=1}^6 \frac{
  L_{\tilde{\chi}_i^0\tilde{u}_k q'}
  L_{\tilde{\chi}_j^+\tilde{u}_k q }^\ast
 }{u_{\tilde{u}_k}} \right],~\nonumber \\
 Q_{LL}^{t0-} &=&
 \frac{1}{x_W\,(1-x_W)} \left[ \frac{L_{Wqq'} O^{R}_{ij}}{s_w}
 - \sum_{k=1}^6 \frac{
  L_{\tilde{\chi}_i^0\tilde{d}_k q }^\ast
  L_{\tilde{\chi}_j^+\tilde{d}_k q'}
 }{t_{\tilde{d}_k}} \right],~\nonumber \\
 Q_{RR}^{u0-} &=&
 \frac{1}{x_W\,(1-x_W)} \left[ \frac{R_{Wqq'} O^{R}_{ij}}{s_w}
 + \sum_{k=1}^6 \frac{
  R_{\tilde{\chi}_i^0\tilde{u}_k q'}
  R_{\tilde{\chi}_j^+\tilde{u}_k q }^\ast
 }{u_{\tilde{u}_k}} \right],~\nonumber
\eea
\bea
 Q_{RR}^{t0-} &=&
 \frac{1}{x_W\,(1-x_W)} \left[ \frac{R_{Wqq'} O^{L}_{ij}}{s_w}
 - \sum_{k=1}^6 \frac{
  R_{\tilde{\chi}_i^0\tilde{d}_k q }^\ast
  R_{\tilde{\chi}_j^+\tilde{d}_k q'}
 }{t_{\tilde{d}_k}} \right],~\nonumber \\
 Q_{LR}^{u0-} &=&
 \frac{1}{x_W\,(1-x_W)}
   \sum_{k=1}^6 \frac{
  R_{\tilde{\chi}_i^0\tilde{u}_k q'}
  L_{\tilde{\chi}_j^+\tilde{u}_k q }^\ast
 }{u_{\tilde{u}_k}}        ,~\nonumber \\
 Q_{LR}^{t0-} &=&
 \frac{1}{x_W\,(1-x_W)}
   \sum_{k=1}^6 \frac{
  L_{\tilde{\chi}_i^0\tilde{d}_k q }^\ast
  R_{\tilde{\chi}_j^+\tilde{d}_k q'}
 }{t_{\tilde{d}_k}}        ,~\nonumber \\
 Q_{RL}^{u0-} &=&
 \frac{1}{x_W\,(1-x_W)}
   \sum_{k=1}^6 \frac{
  L_{\tilde{\chi}_i^0\tilde{u}_k q'}
  R_{\tilde{\chi}_j^+\tilde{u}_k q }^\ast
 }{u_{\tilde{u}_k}}        ,~\nonumber \\
 Q_{RL}^{t0-} &=&
 \frac{1}{x_W\,(1-x_W)}
   \sum_{k=1}^6 \frac{
  R_{\tilde{\chi}_i^0\tilde{d}_k q }^\ast
  L_{\tilde{\chi}_j^+\tilde{d}_k q'}
 }{t_{\tilde{d}_k}}.
 \label{eq:9}
\eea
The related charge-conjugate process $q\bar{q}'\to\tilde{\chi}^+_i
\tilde{\chi}^0_j$ \cite{Bozzi:2007me} is obtained by complex conjugation
and making the replacements $i\leftrightarrow j$, $q\leftrightarrow q'$,
and $LR\leftrightarrow RL$ in these charges. In the case of nonmixing
squarks with neglected Yukawa couplings, we agree with the results of Ref.\
\cite{Beenakker:1999xh}, provided we correct a sign in their Eq.\ (2) as
described in the Erratum.

Finally, the charges for neutralino-pair production are given by
\bea
 Q_{LL}^{u00} &=&
 \frac{1}{x_W\,(1-x_W)\,\sqrt{1+ \delta_{ij}}} \left[ \frac{L_{Zqq'}
 O^{\prime\prime L}_{ij} }{s_z} + \sum_{k=1}^6
 \frac{L_{\tilde{\chi}_i^0\tilde{q}_k q^\prime } L_{\tilde{\chi}_j^0
 \tilde{q}_k q}^\ast}{u_{\tilde{q}_k}} \right] ,~ \nonumber \\
 Q_{LL}^{t00} &=& \frac{1}{x_W\,(1-x_W)\,\sqrt{1+ \delta_{ij}}}
 \left[ \frac{L_{Zqq'} O^{\prime\prime R}_{ij} }{s_z} -
 \sum_{k=1}^6 \frac{L_{\tilde{\chi}_i^0\tilde{q}_k q }^\ast
 L_{\tilde{\chi}_j^0\tilde{q}_k q^\prime }}{t_{\tilde{q}_k}}
 \right] ,~ \nonumber \\
 Q_{RR}^{u00} &=&
 \frac{1}{x_W\,(1-x_W)\,\sqrt{1+ \delta_{ij}}} \left[ \frac{R_{Zqq'}
 O^{\prime\prime R}_{ij} }{s_z} + \sum_{k=1}^6
 \frac{R_{\tilde{\chi}_i^0\tilde{q}_k q^\prime } R_{\tilde{\chi}_j^0
 \tilde{q}_k q}^\ast}{u_{\tilde{q}_k}} \right] ,~ \nonumber \\
 Q_{RR}^{t00} &=& \frac{1}{x_W\,(1-x_W)\,\sqrt{1+ \delta_{ij}}}
 \left[ \frac{R_{Zqq'} O^{\prime\prime L}_{ij} }{s_z} -
 \sum_{k=1}^6 \frac{R_{\tilde{\chi}_i^0\tilde{q}_k q }^\ast
 R_{\tilde{\chi}_j^0\tilde{q}_k q^\prime }}{t_{\tilde{q}_k}}
 \right] ,~\nonumber \\
 Q_{LR}^{u00} &=&
 \frac{1}{x_W\,(1-x_W)\,\sqrt{1+ \delta_{ij}}} \sum_{k=1}^6
 \frac{R_{\tilde{\chi}_i^0\tilde{q}_k q^\prime } L_{\tilde{\chi}_j^0
 \tilde{q}_k q}^\ast}{u_{\tilde{q}_k}} ,~\nonumber \\
 Q_{LR}^{t00} &=& \frac{1}{x_W\,(1-x_W)\,\sqrt{1+ \delta_{ij}}}
 \sum_{k=1}^6 \frac{L_{\tilde{\chi}_i^0\tilde{q}_k q }^\ast
 R_{\tilde{\chi}_j^0\tilde{q}_k q^\prime }}{t_{\tilde{q}_k}} ,~
 \nonumber \\
 Q_{RL}^{u00} &=& \frac{1}{x_W\,(1-x_W)\,\sqrt{1+
 \delta_{ij}}} \sum_{k=1}^6 \frac{L_{\tilde{\chi}_i^0\tilde{q}_k q^\prime
 } R_{\tilde{\chi}_j^0\tilde{q}_k q}^\ast}{u_{\tilde{q}_k}} ,~ \nonumber \\
 Q_{RL}^{t00} &=& \frac{1}{x_W\,(1-x_W)\,\sqrt{1+ \delta_{ij}}}
 \sum_{k=1}^6 \frac{R_{\tilde{\chi}_i^0\tilde{q}_k q }^\ast
 L_{\tilde{\chi}_j^0\tilde{q}_k q^\prime }}{t_{\tilde{q}_k}},
 \label{eq:10}
\eea
which agrees with the results of Ref.\ \cite{Bozzi:2007me} and also with
those of Ref.\ \cite{Gounaris:2004fm} in the case of nonmixing squarks.

\section{Numerical results}
\label{sec:3}

We now present numerical predictions for the cross sections and single- and
double-spin asymmetries of gaugino-pair production at the polarized $pp$
collider RHIC \cite{rhicspin} and possible polarization upgrades of the
$p\bar{p}$ and $pp$ colliders Tevatron \cite{Baiod:1995eu} and LHC
\cite{roeck}. Thanks to the QCD factorization theorem, total unpolarized
hadronic cross sections
\bea
 \sigma &~=&
 \int_{4m^2/S}^1\!\d\tau\!\!
 \int_{-1/2\ln\tau}^{1/2\ln\tau}\!\!\d y
 \int_{t_{\min}}^{t_{\max}} \d t \
 f_{a/A}(x_a,\mu_F) \ f_{b/B}(x_b,\mu_F) \ {\d\hat{\sigma}\over\d t}
\eea
can be calculated by convolving the relevant partonic cross sections
d$\hat{\sigma}$/d$t$, computed in Sec.\ \ref{sec:2}, with universal parton
densities $f_{a/A}$ and $f_{b/B}$ of partons $a,b$ in the hadrons $A,B$,
which depend on the longitudinal momentum fractions of the two partons
$x_{a,b} = \sqrt{\tau}e^{\pm y}$ and on the unphysical factorization scale
$\mu_F$. Polarized cross sections are computed similarly by replacing
either one or both of the unpolarized parton densities $f_{a,b}(x_{a,b},
\mu_F)$ with their polarized equivalents $\Delta f_{a,b}(x_{a,b},\mu_F)$
and the unpolarized partonic cross section $\d\hat{\sigma}$, given in
Eq.\ (\ref{eq:5}), with its single- or double-polarized equivalent given
in Eq.\ (\ref{eq:6}).

Efforts over the past three decades have produced extensive data sets for
polarized deep-inelastic scattering (DIS), resulting in a good knowledge
in particular of the polarized valence-quark (non-singlet) distributions.
For consistency with our leading order (LO) QCD calculation in the collinear
approximation, where all squared quark masses (except for the top-quark
mass) $m_q^2\ll s$, we employ related sets of unpolarized
\cite{Gluck:1998xa} and polarized \cite{Gluck:2000dy} LO parton densities.
We estimate the theoretical uncertainty due to the less well known polarized
parton densities by showing our numerical predictions for both the GRSV2000
LO standard (STD) and valence (VAL) parameterizations, which treat the polarized
sea-quarks in a flavor-symmetric or flavor-asymmetric way. The polarized gluon
density could not be constrained very well in the fits to the DIS data, but it
fortunately does not enter directly in our analysis.

Results from semi-inclusive DIS with an identified hadron in the final state
have the promise to put individual constraints on the various quark flavor
distributions in the nucleon. In addition, precise asymmetry measurements from
RHIC are expected to put significant constraints on the polarized gluon
distribution. A first step in this direction has been undertaken very recently by
including semi-inclusive DIS data from the SMC, HERMES and COMPASS experiments
and $\pi^0$ and jet production data from the PHENIX and STAR collaborations in a
global analysis \cite{deFlorian:2008mr}.

If not stated
otherwise, we set the factorization scale $\mu_F$ to the average mass of the
final state SUSY particles. The bottom- and top-quark densities in the
proton are small and absent in the GRV and GRSV parameterizations, as is the
charm-quark density. We therefore consider for squark exchanges only the
SUSY-partners of the light quark flavors without mixing and all degenerate
in mass. The corresponding uncertainty is estimated by giving predictions
for two different squark masses, one at the mass limit set by the D0
collaboration at 325 GeV \cite{Abazov:2006bj} and one for a typical
SUSY-breaking scale of 1 TeV.

\subsection{Gaugino masses and mixings}

We wish to study the correlations of beam polarizations and the
gaugino/Higgsino fractions of charginos and neutralinos without referring to
a particular SUSY-breaking model. Furthermore, we wish to keep the physical
gaugino masses as constant as possible, since the absolute cross sections
depend strongly on them through trivial phase space effects. We start
therefore by fixing the lightest chargino mass $m_{\tilde{\chi}^\pm_1}$ to
either 80 GeV (for our RHIC predictions) or 151 GeV (for our Tevatron and
LHC predictions). The relatively strong limit of 151 GeV has recently been
obtained by the CDF collaboration at Run II of the Tevatron and holds for a
constrained MSSM with light nonmixing sleptons \cite{:2007nz}. On the other
hand, charginos with a mass as low as 80 GeV may still be allowed, if
they are gauginolike, their
mass difference with the lightest neutralino is very small, and if the
sneutrinos are light \cite{Abbiendi:2002vz,Abdallah:2003xe}. The
second-lightest neutralino usually stays close in mass to the lightest
chargino (see below) and must be heavier than 62.4 GeV, while the lightest
neutralino can be half as heavy and is constrained to masses above 32.5-46
GeV, depending again on the sfermion masses \cite{Yao:2006px}. The
associated production of the second-lightest neutralino with the lightest
chargino is usually experimentally easily identifiable through the
gold-plated tri-lepton decay. It has been pointed out that the
electroweak precision fits improve when including heavy sfermions as proposed
by split-SUSY scenarios, but light gauginos or Higgsinos with masses close to
the current exclusion limits \cite{Martin:2004id}.

In the MSSM, the gaugino masses and mixings depend on the {\em a priori}
unknown SUSY-breaking parameters $M_1$, $M_2$, $\mu$, and on $\tan\beta$
(see App.\ \ref{sec:a}). Taking $\tan\beta=10$ and assuming $B$-ino and $W$-ino
mass unification at the GUT scale, so that $M_1={5\over3}\tan^2\theta_W M_2
\simeq0.5 M_2$ at the electroweak scale, we can compute the Higgsino mass
parameter $\mu$ from Eq.\ (\ref{eq:a14}),
\bea
 \mu&=&\frac{m_W^2~M_2~s_{2\beta}\pm m_{\tilde{\chi}^\pm_1}
 \sqrt{\left(m_{\tilde{\chi}^\pm_1}^2-M_2^2-m_W^2\right)^2-
 m_W^4~c^2_{2\beta}}}{M_2^2-m_{\tilde{\chi}^\pm_1}^2},
\eea
as a function of the only remaining parameter $M_2$, once the lightest
chargino mass $m_{\tilde{\chi}^\pm_1}$ is fixed. Since the one-loop
contribution to the anomalous magnetic moment $a_\mu=(g_\mu-2)/2$ of the
muon induced by gauginos and sleptons of common mass $M_{\rm SUSY}$ is
approximately given by \cite{Moroi:1995yh}
\bea
 a_\mu^{\rm SUSY,~1-loop} &=&
 13\times 10^{-10} \lr{100~{\rm GeV}\over M_{\rm SUSY}}\rr^2\tan\beta \
 {\rm sgn}(\mu),
\eea
negative values of $\mu$ would increase, not decrease, the disagreement
between the recent BNL measurement and the theoretical SM value of $a_\mu$
\cite{Yao:2006px}. The region $\mu<0$ is therefore disfavored, and
we take $\mu>0$ unless noted otherwise.

As the off-diagonal matrix elements of the gaugino mass matrices depend on
$\sin\beta$ and $\cos\beta$ (see App.\ \ref{sec:a}), one might be
tempted to fix $M_2$, e.g.\ to $2\,m_{\tilde{\chi}^\pm_1}$, and study rather
the variation of the chargino/neutralino masses and gaugino/Higgsino
fractions with $\tan\beta$. However, this parameter can often be constrained from
the Higgs sector alone \cite{Buescher:2005re}, at least if it is large
\cite{Gunion:1996cn}; otherwise measurements from the sfermion or neutralino
sector may still be necessary \cite{Denegri:1999pe}. Furthermore, $\sin\beta$
and $\cos\beta$ vary significantly only for low $\tan\beta=2\,...\,10$.
In this range, the gaugino fraction of the lightest negative chargino
decreases, e.g., from 40\% to 20\% in the optimal case of
$M_2=2\,m_{\tilde{\chi}^\pm_1}=160$ GeV.

\begin{figure}
 \centering
 \epsfig{file=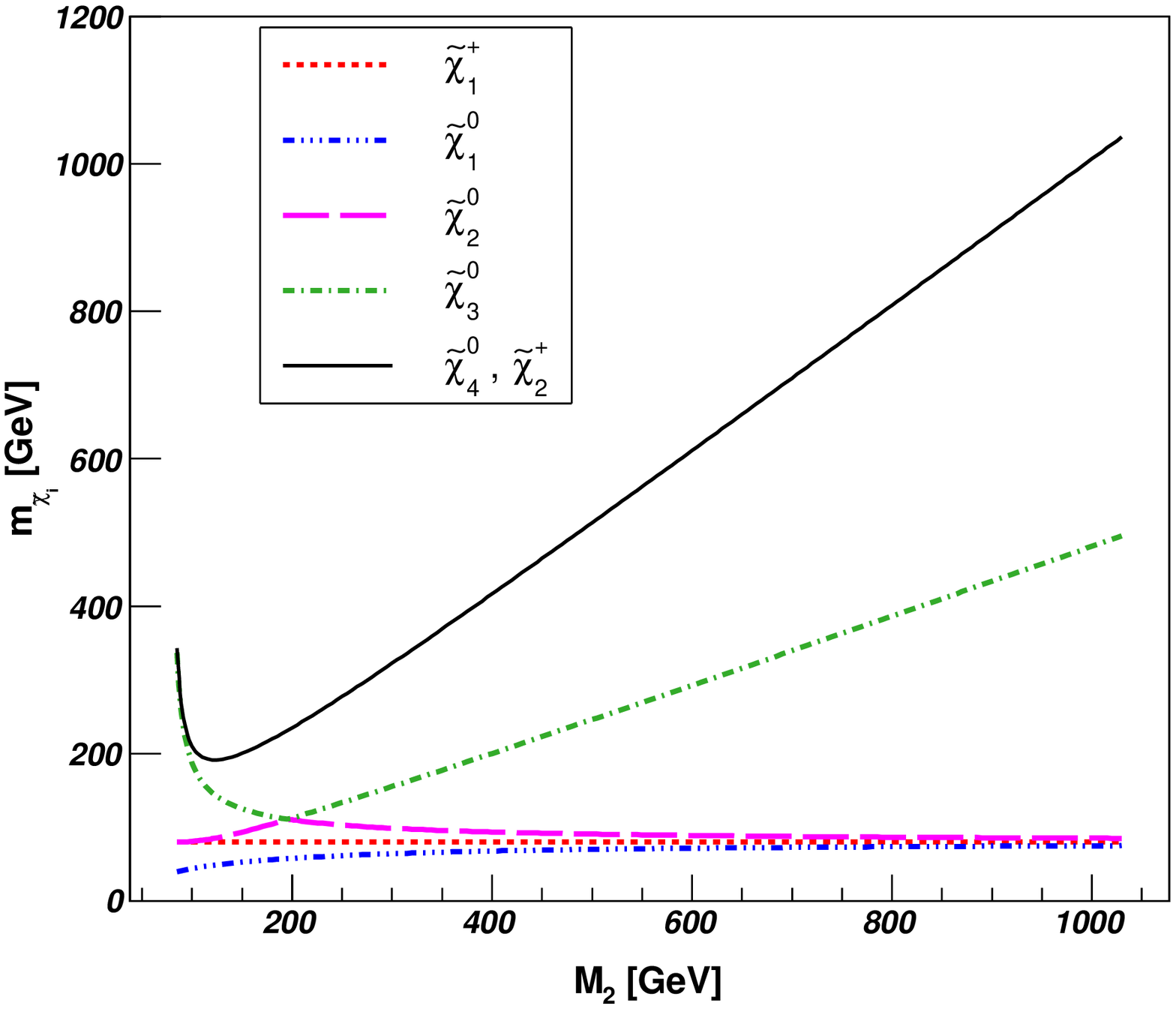,width=.49\columnwidth,clip=}
 \epsfig{file=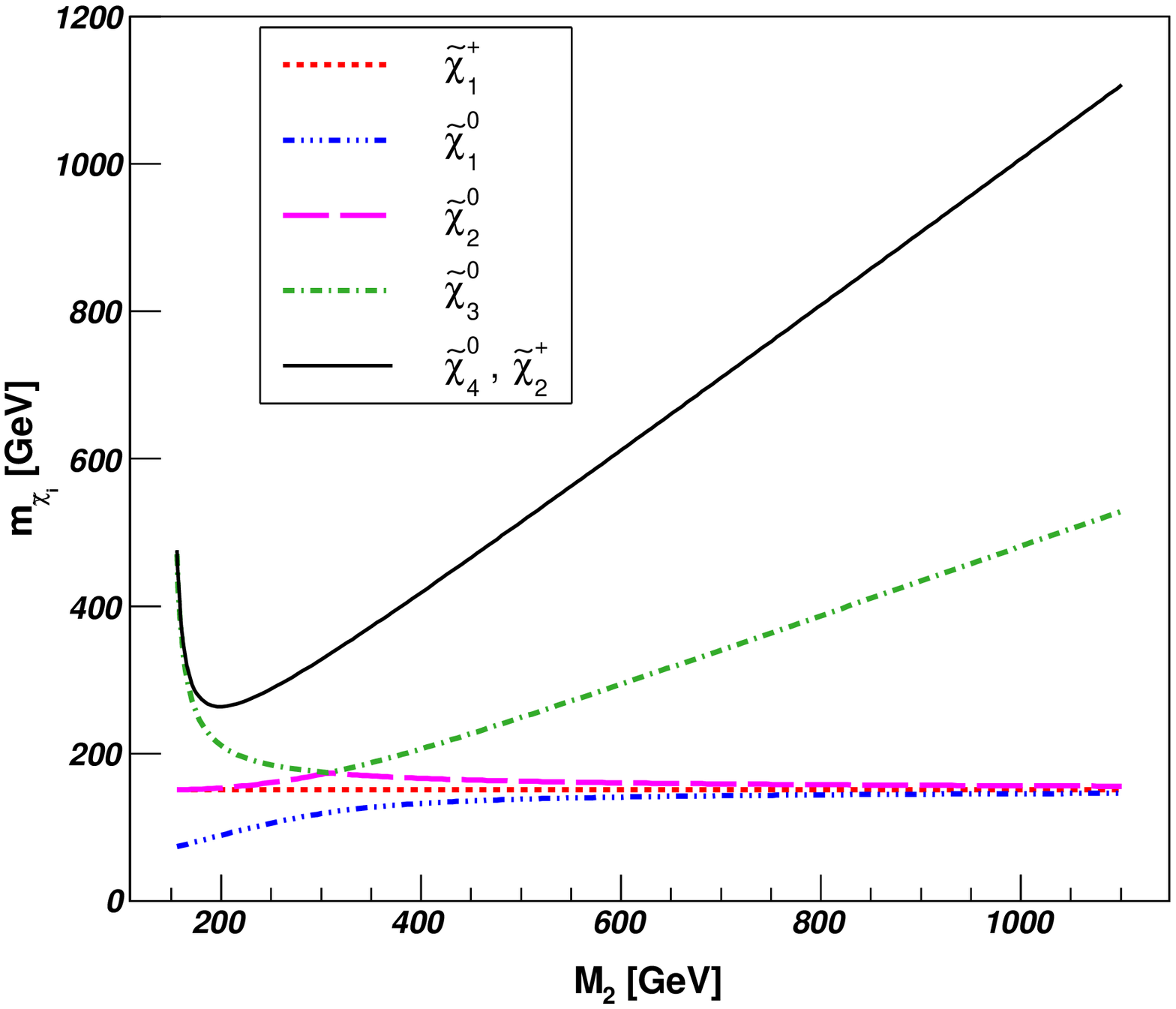,width=.49\columnwidth,clip=}
 \caption{\label{fig:2}Neutralino and chargino masses as a function of the
          SUSY-breaking parameter $M_2$ for a fixed lightest chargino mass of
          $m_{\tilde{\chi}_1}^\pm=80$ GeV (left) and 151 GeV (right). We choose
          $\tan\beta=10$, $\mu>0$ using Eq.\ (12), and fix $M_1={5\over3}\tan^2
          \theta_W M_2$.}
\end{figure}

\begin{figure}
 \centering
 \epsfig{file=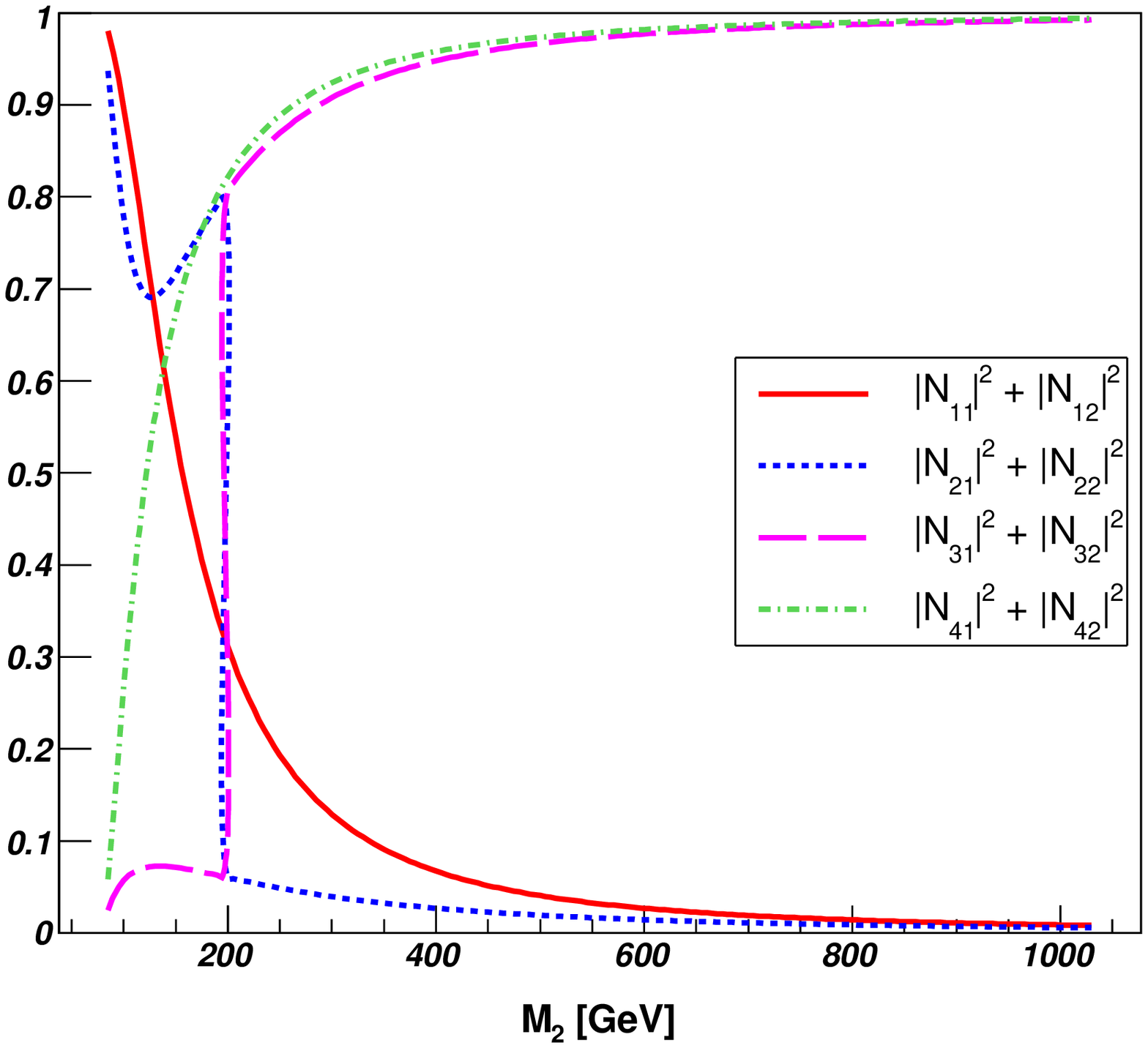,width=.49\columnwidth,clip=}
 \epsfig{file=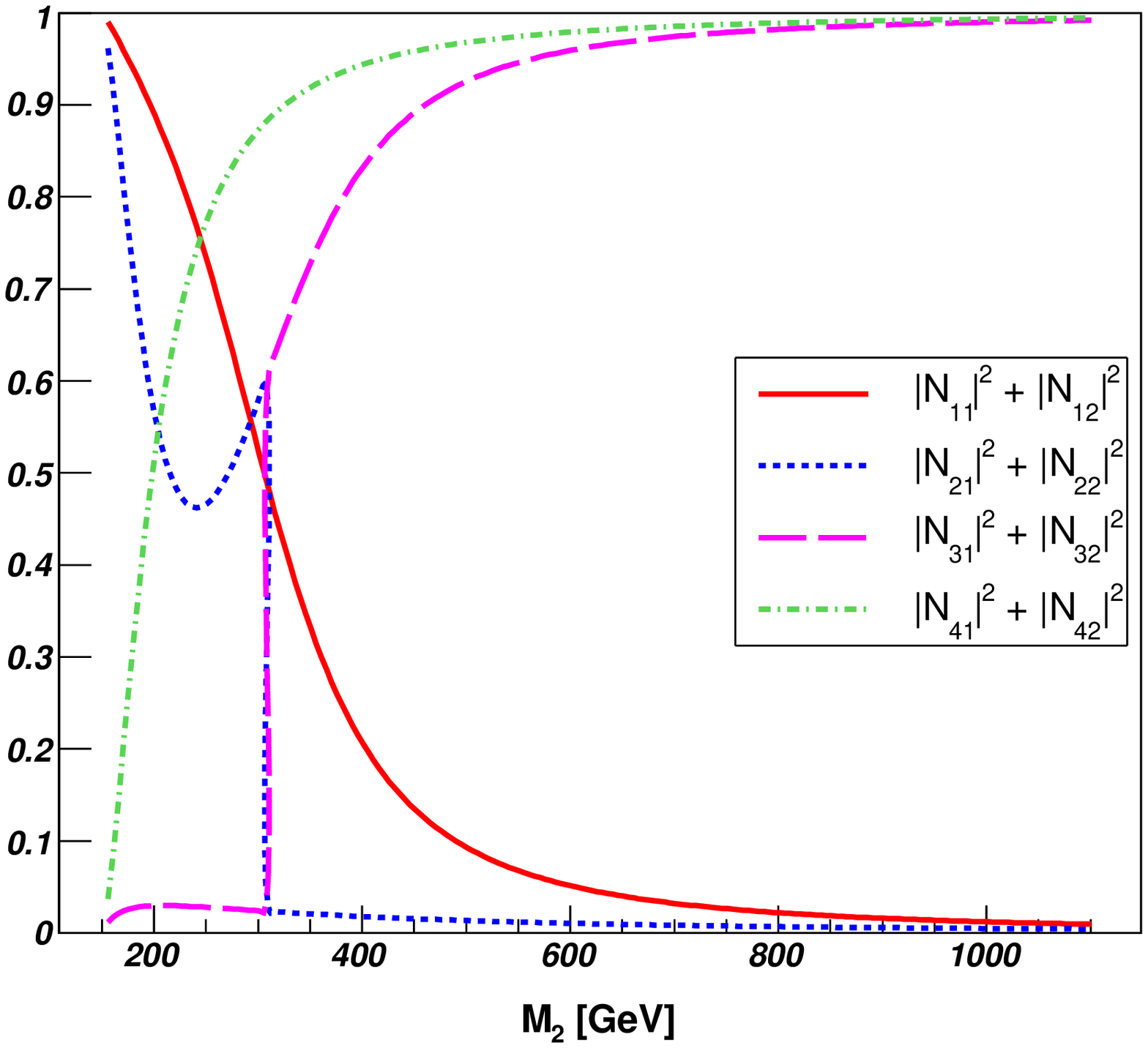,width=.49\columnwidth,clip=}
 \caption{\label{fig:3}Gaugino and Higgsino fractions of the four
          neutralinos as a function of the SUSY-breaking parameter $M_2$ for
          a fixed lightest chargino mass of $m_{\tilde{\chi}_1}^\pm=80$ GeV
          (left) and 151 GeV (right).  We choose
          $\tan\beta=10$, $\mu>0$ using Eq.\ (12), and fix $M_1={5\over3}\tan^2
          \theta_W M_2$.}
\end{figure}

\begin{figure}
 \centering
 \epsfig{file=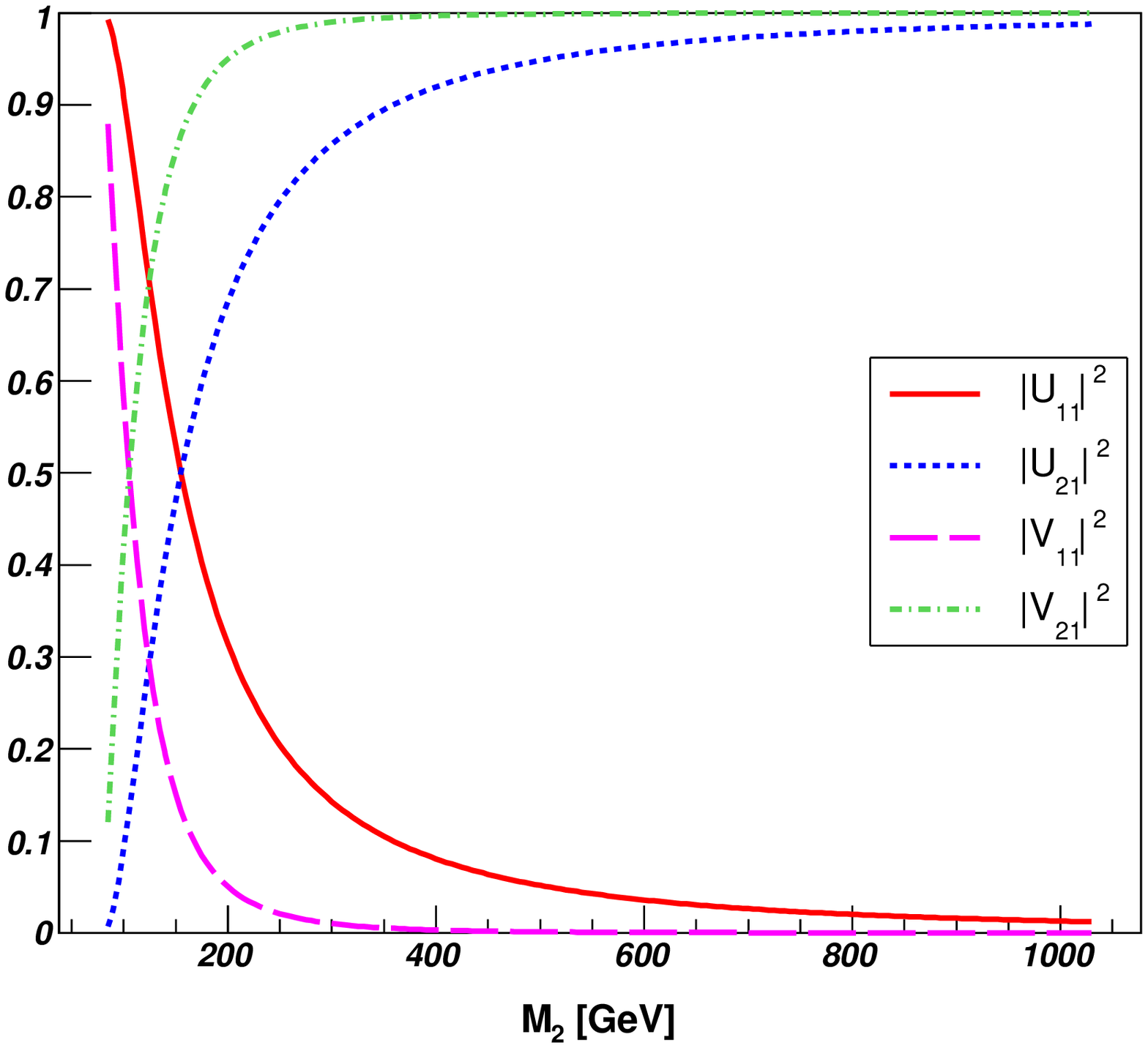,width=.49\columnwidth,clip=}
 \epsfig{file=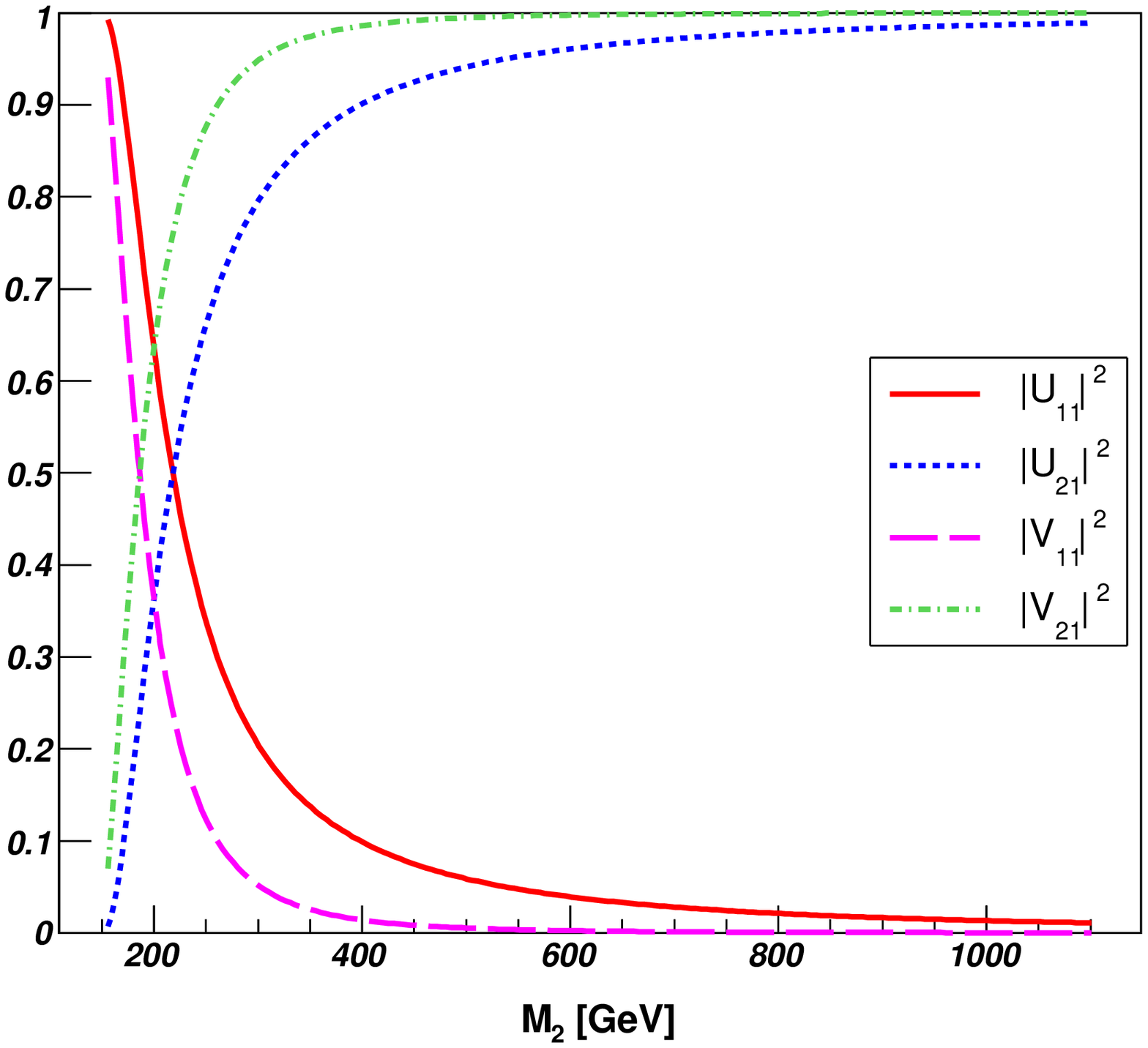,width=.49\columnwidth,clip=}
 \caption{\label{fig:4}Gaugino and Higgsino fractions for charginos as a
          function of the SUSY-breaking parameter $M_2$ for a fixed lightest
          chargino mass of $m_{\tilde{\chi}_1}^\pm=80$ GeV (left) and 151
          GeV (right).  We choose
          $\tan\beta=10$, $\mu>0$ using Eq.\ (12), and fix $M_1={5\over3}\tan^2
          \theta_W M_2$.}
\end{figure}

In Fig.\ \ref{fig:2} we show the physical masses of the two charginos and
%
%
the four neutralinos as a function of $M_2$ for $m_{\tilde{\chi}^\pm_1}=80$
GeV (left) and 151 GeV (right). The lightest chargino mass (short-dashed
line) is, of course, constant in both cases. As mentioned above, the mass of
the second-lightest neutralino stays close to it, except around $M_2=190$
GeV (320 GeV), where an avoided crossing with $m_{\tilde{\chi}^0_3}$ occurs,
which is typical of Hermitian matrices depending continuously on a single
parameter. At this point, these two neutralino eigenstates change character,
as can clearly be seen from the gaugino fractions plotted in Fig.\
\ref{fig:3}.
%
%
While for small values of $M_2\ll|\mu|$ the lighter neutralinos,
diagonalized by the matrix $N$, are gauginolike, they become Higgsino-like
for large values of $M_2\gg|\mu|$. Furthermore, in this region the mass
difference between the lightest neutralino and chargino becomes small (see
Fig.\ \ref{fig:2}). It can also be seen from this figure that the heavier
chargino and the heaviest neutralino are mass-degenerate for all values of
$M_2$ and that their mass grows linearly with $M_2$, when $M_2\gg|\mu|$. The
gaugino fractions of the negative and positive charginos, diagonalized by
the matrices $U$ and $V$, are shown in Fig.\ \ref{fig:4}. They behave
%
%
similarly to those of the lightest and heaviest neutralinos. We will
frequently refer to these well-known variations of the neutralino/chargino
masses and gaugino/Higgsino fractions in the subsequent sections when discussing
the behavior of cross sections and asymmetries.

\subsection{RHIC cross sections and asymmetries}

RHIC is scheduled to operate in the years 2009 through 2012 in its polarized
$pp$ mode at an increased center-of-mass energy of $\sqrt{S}=500$ GeV and with
a large integrated luminosity of 266 pb$^{-1}$ during each of the ten-week
physics runs \cite{rhicspin}. It has been demonstrated that polarization loss
during RHIC beam acceleration and storage can be kept small, so that a
polarization degree of about 45\% has already been and 65\%-70\% may ultimately be
reached \cite{Roser:2008zz}. It is therefore interesting to investigate the
influence of proton beam polarization on production cross sections and
longitudinal spin asymmetries for SUSY particle production at the existing
polarized $pp$ collider RHIC.

\begin{figure}
 \centering
 \epsfig{file=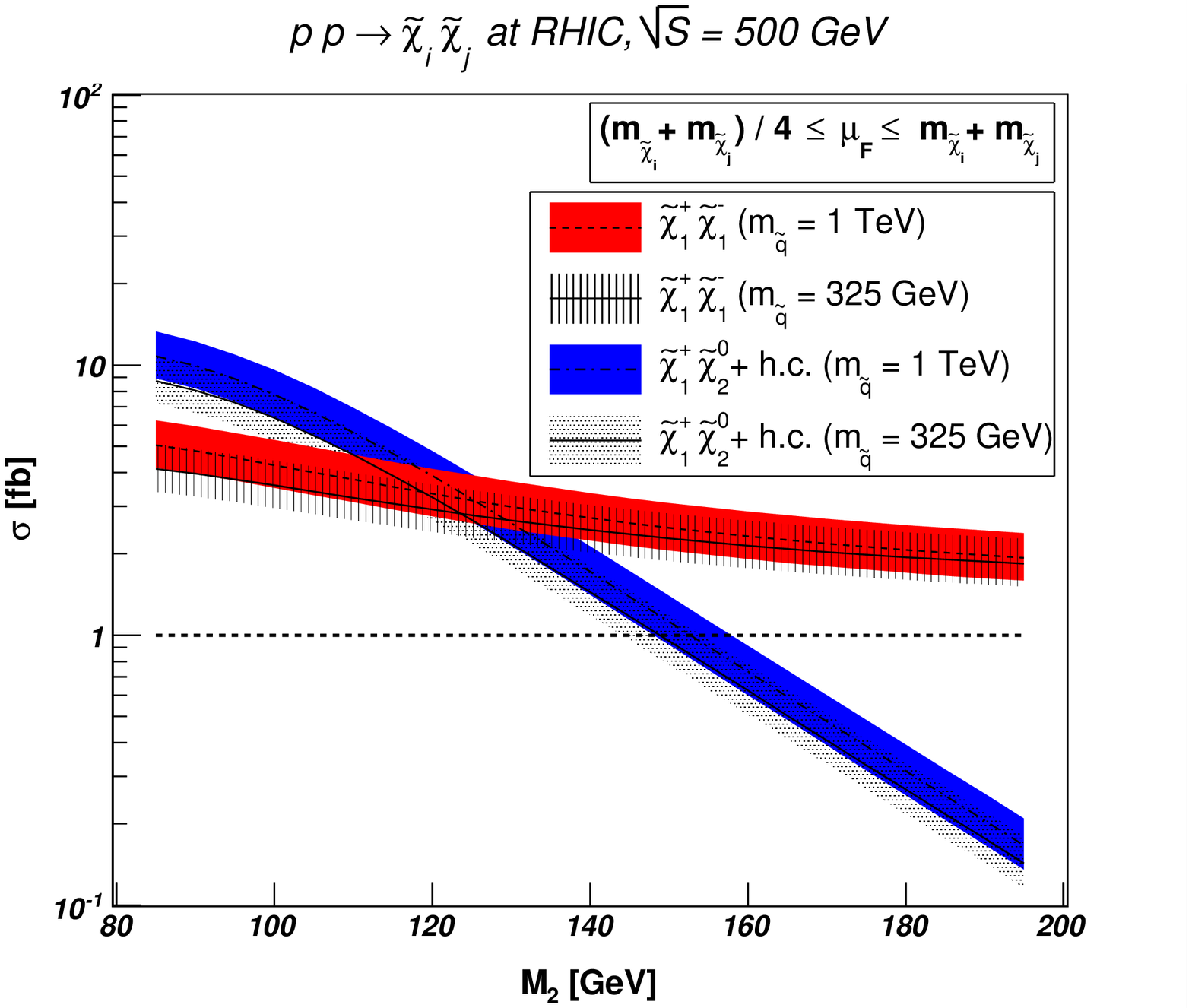,width=.49\columnwidth,clip=}
 \epsfig{file=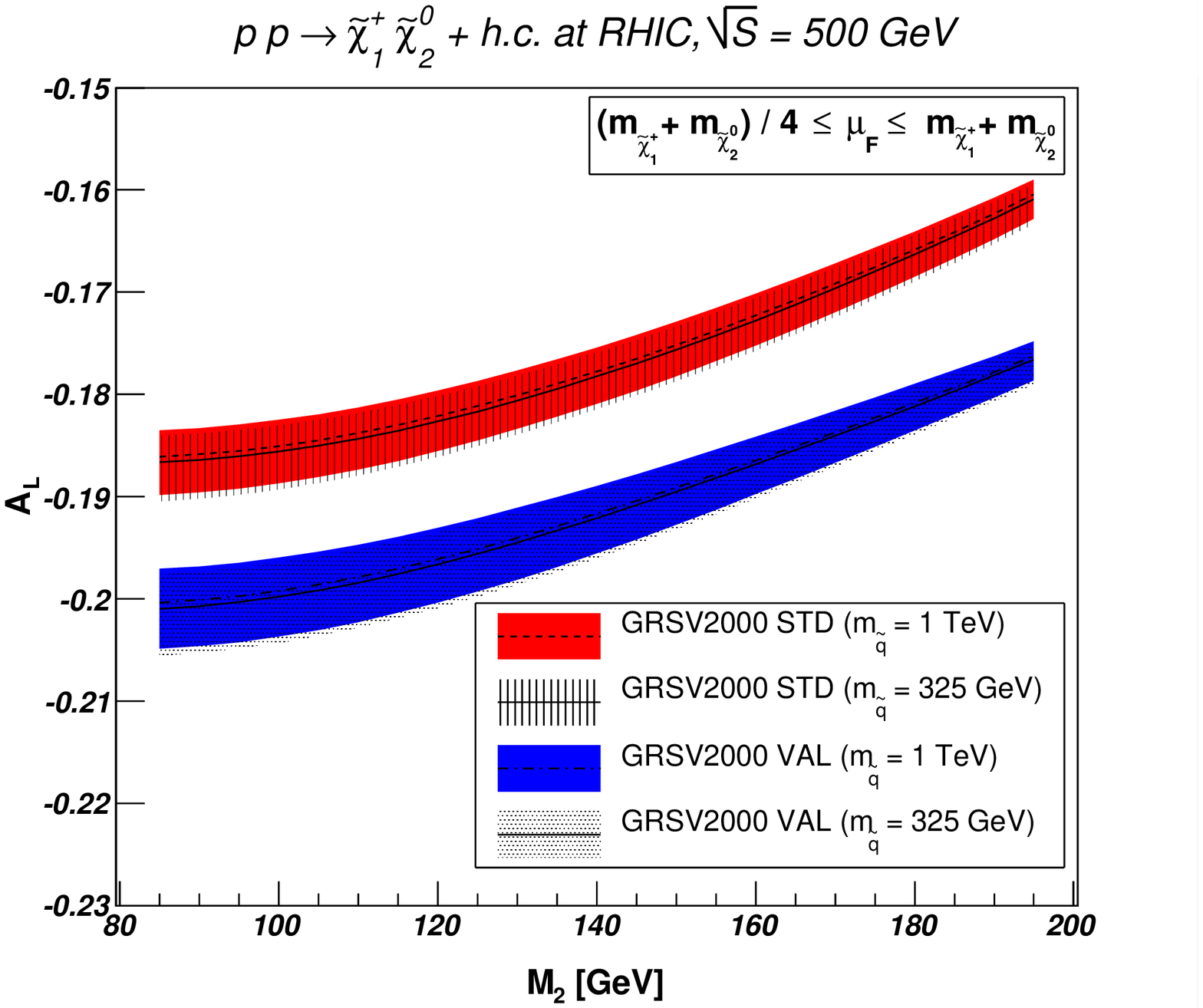,width=.49\columnwidth,clip=}
 \caption{\label{fig:5}Unpolarized gaugino-pair production cross sections
          (left) and single-spin
          asymmetries for chargino-neutralino associated production (right)
          with $m_{\tilde{\chi}^0_2}\simeq m_{\tilde{\chi}^\pm_1}=80$ GeV in
          $pp$ collisions at RHIC and $\sqrt{S}=500$ GeV using LO GRV
          \cite{Gluck:1998xa} and GRSV \cite{Gluck:2000dy} parton densities.
          We choose
          $\tan\beta=10$, $\mu>0$ using Eq.\ (12), and fix $M_1={5\over3}\tan^2
          \theta_W M_2$.}
\end{figure}

In the left part of Fig.\ \ref{fig:5}, we show the total unpolarized cross
%
%
section for the pair production of the lightest chargino of mass 80 GeV
(short-dashed line) and the one for its associated production with the
second-lightest neutralino (dot-dashed line) at the $pp$ collider RHIC,
expected to 
produce a total integrated luminosity of about 1 fb$^{-1}$ during the next
four years \cite{rhicspin}. Both cross sections exceed 1 fb (short-dashed
horizontal line, corresponding to one produced event) in most of the
$M_2$ range shown and depend little on the squark mass, indicating that
$s$-channel gauge-boson exchanges dominate. From Eqs.\ (\ref{eq:8}) and
(\ref{eq:9}) and App.\ \ref{sec:b} we learn indeed that, in the absence of
heavy bottom- and top-quarks, squark exchanges contribute only to
$Q_{LL}^{t,u}$ for
chargino pairs and in addition to $Q_{LR}^u$ and $Q_{RL}^t$ for the
associated channel. For the latter, we sum both charge conjugate processes,
even though it might be interesting to identify the chargino charge, given
that the dependence of its gaugino fraction on $M_2$ is slightly different
for the two charges (see Fig.\ \ref{fig:4}). The pair production of the
second-lightest neutralino (not shown) does receive squark contributions
from all generalized charges, but the corresponding cross section lies below
10$^{-2}$ fb and is therefore invisible at RHIC. As our cross sections are
computed at LO, they depend to some extent on the factorization scale
$\mu_F$. Since this scale is unphysical and unknown, we vary it in the
traditional way by a factor of two around the average final state
mass, representing the large perturbative scale in the partonic cross
section (shaded bands).

Among the bosons exchanged in the $s$-channel, the $W$-boson is most
sensitive to the polarization of the initial quarks and antiquarks, and
consequently the single-spin asymmetry for the associated channel, shown in
the right part of Fig.\ \ref{fig:5}, reaches large values of around -20\%.
Note that polarization of the proton beam(s) will not be perfect, so that
all calculated single-spin (double-spin) asymmetries should be multiplied by
the degree of beam polarization $P_{A,B} \simeq 0.7$ (squared).
 This follows from the fact that, while the analytical
counting rates $N$ for realistic polarizations are in principle rather complex,
e.g.\
\bea
N_{1,1}&=&{1\over2} (1+P_A) {1\over2} (1+P_B) \sigma_{1,1} + {1\over2} (1+P_A) {1\over2} (1-P_B) \sigma_{1,-1}
\nonumber \\
 &+&     {1\over2} (1-P_A) {1\over2} (1+P_B) \sigma_{-1,1} + {1\over2} (1-P_A) {1\over2} (1-P_B) \sigma_{-1,-1},
\eea
where $\sigma_{i,j}$ mean the ideal cross sections and $P_A$ and $P_B$ the beam
polarizations, the double-spin asymmetry
\bea
 A_{LL} &=& {N_{1,1} + N_{-1,-1} - N_{1,-1} - N_{-1,1}\over
        N_{1,1} + N_{-1,-1} + N_{1,-1} + N_{-1,1}}
\eea
simplifies to give $A_{LL}~=~P_A~P_B~A'_{LL}$
with $A'_{LL}$ composed of $\sigma_{i,j}$. The same applies to the single-spin
asymmetry $A_L$ with just multiplication by $P_A$ (or $P_B$).

As the mass of the neutralino increases and the gaugino fractions of the
chargino and neutralino fall up to $M_2\leq200$ GeV, the cross
section and the absolute value of the asymmetry decrease, too.
For these values of $M_2$, the conditions of the LEP chargino mass limit
still apply.
The uncertainty in the scale variation is with 0.5\% considerably smaller than the
variation in the asymmetry of 2\%, while the uncertainty coming from the polarized
parton densities is with 1.5\% of almost comparable size. Single-spin
asymmetry measurements for associated chargino-neutralino production at the
only existing polarized hadron collider RHIC could therefore be used to
determine the gaugino and Higgsino components of charginos and neutralinos,
provided the polarized quark and antiquark densities at momentum fractions
of $x_{a,b}\simeq2\times80$ GeV / 500 GeV $=0.32$ are slightly better
constrained. For the double-spin asymmetry (not shown), the parton density
uncertainty exceeds the variation and leads to a sign change of the
relatively small asymmetry (+6\%/-3\%), so that in this case no useful
information on the gaugino/Higgsino mixing can be extracted.

The single- and double-spin asymmetries for neutralino pairs reach similar
sizes as those for the associated channel, since the left- and right-handed
couplings of the $Z$-boson exchanged in the $s$-channel are also different.
However, we do not show them here, since the corresponding cross section is
unfortunately too small at RHIC, as mentioned above. The variation of the
asymmetries would, indeed, be quite dramatic: $A_L$ changes its sign from
-20\% to +20\% for $M_2\leq200$ GeV, and $A_{LL}$ falls from -5\% to
-20\%.

For chargino pairs, massless photons can be exchanged in the $s$-channel,
see Eq.\ (\ref{eq:8}), which leads to single- and double-spin asymmetries
(not shown) that vary very little with $M_2$ and that can therefore not be
used to extract information on gaugino/Higgsino mixing. In addition, these
asymmetries depend strongly on the polarized parton densities.

\subsection{Tevatron cross sections and asymmetries}

The $p\bar{p}$ collider Tevatron
will continue running in 2009 and possibly in 2010, and
the future accelerator program at Fermilab is currently less clear than ever.
The feasibility of polarizing the proton beam has been demonstrated many years
ago \cite{Baiod:1995eu}. It would require replacing some of the dipoles with
higher-field magnets to gain space to install the six required Siberian snakes
at a very moderate cost \cite{Krisch:1998zm} and would represent an interesting
possibility for QCD studies as well as new physics searches. Given the recent
impressive achievements at RHIC, the degree of polarization should be comparable,
i.e.\ about 65\%-70\%. Polarization of the antiproton beam is, however, much more
challenging.

\begin{figure}
 \centering
 \epsfig{file=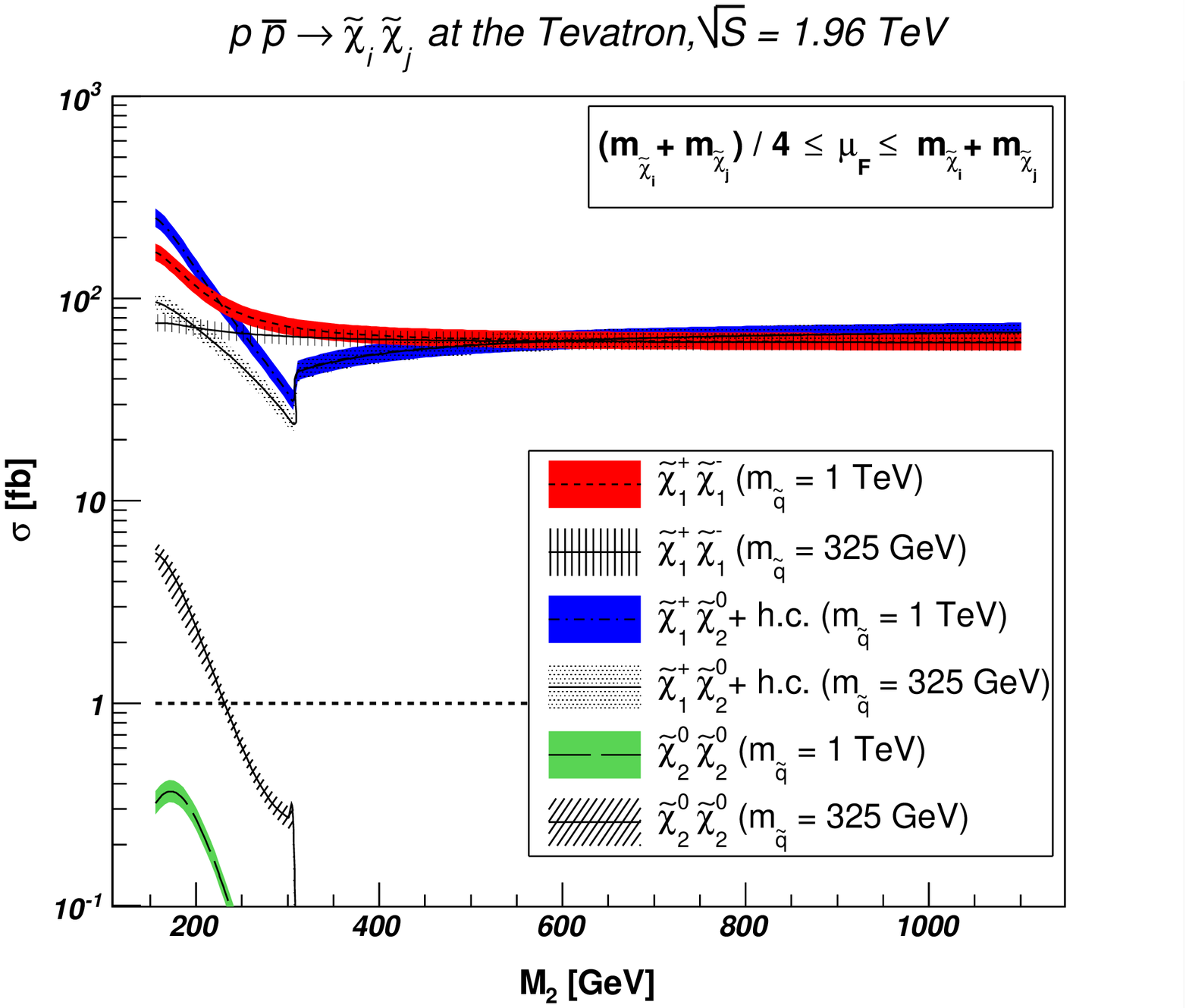,width=.49\columnwidth,clip=}
 \epsfig{file=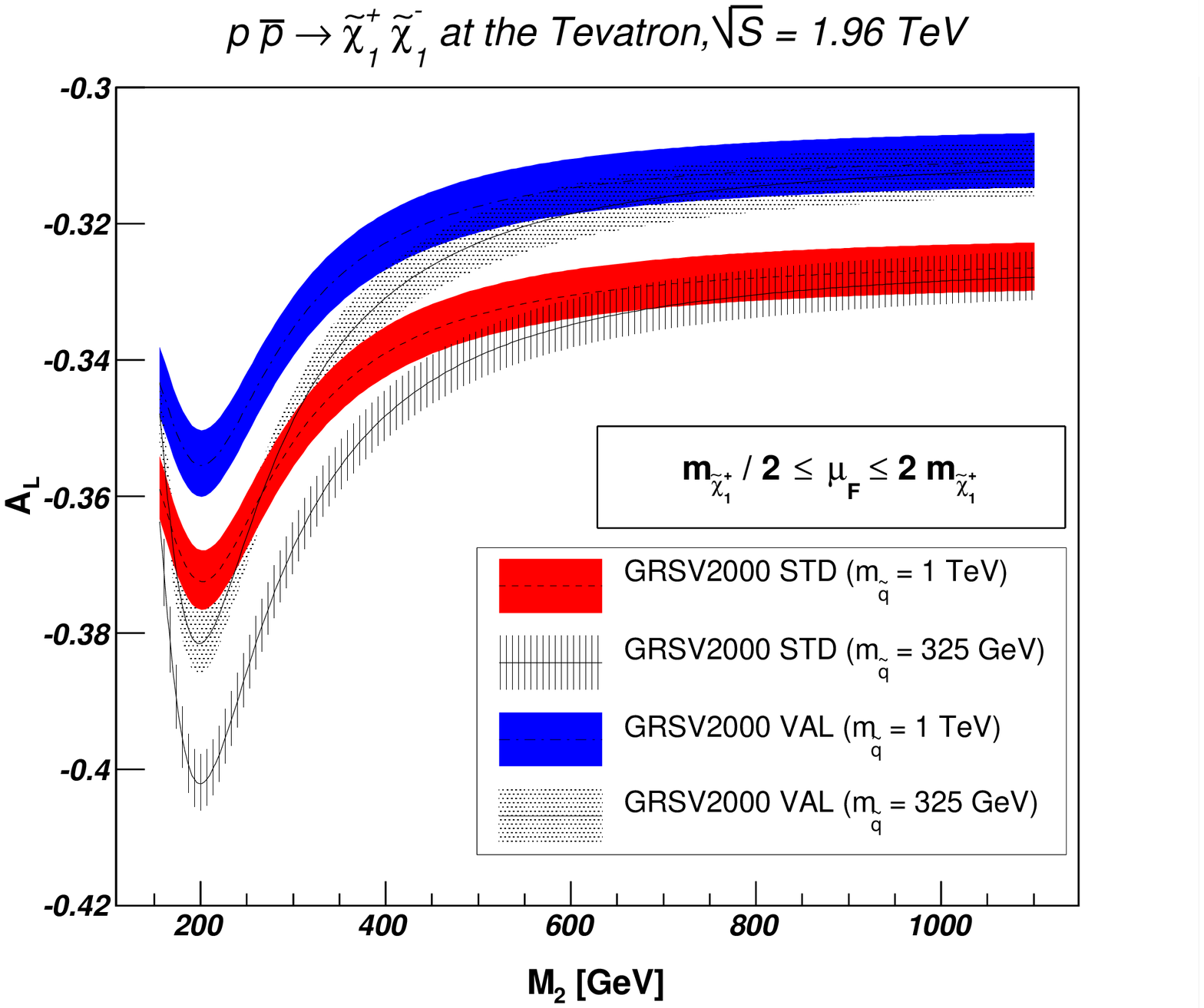,width=.49\columnwidth,clip=}
 \epsfig{file=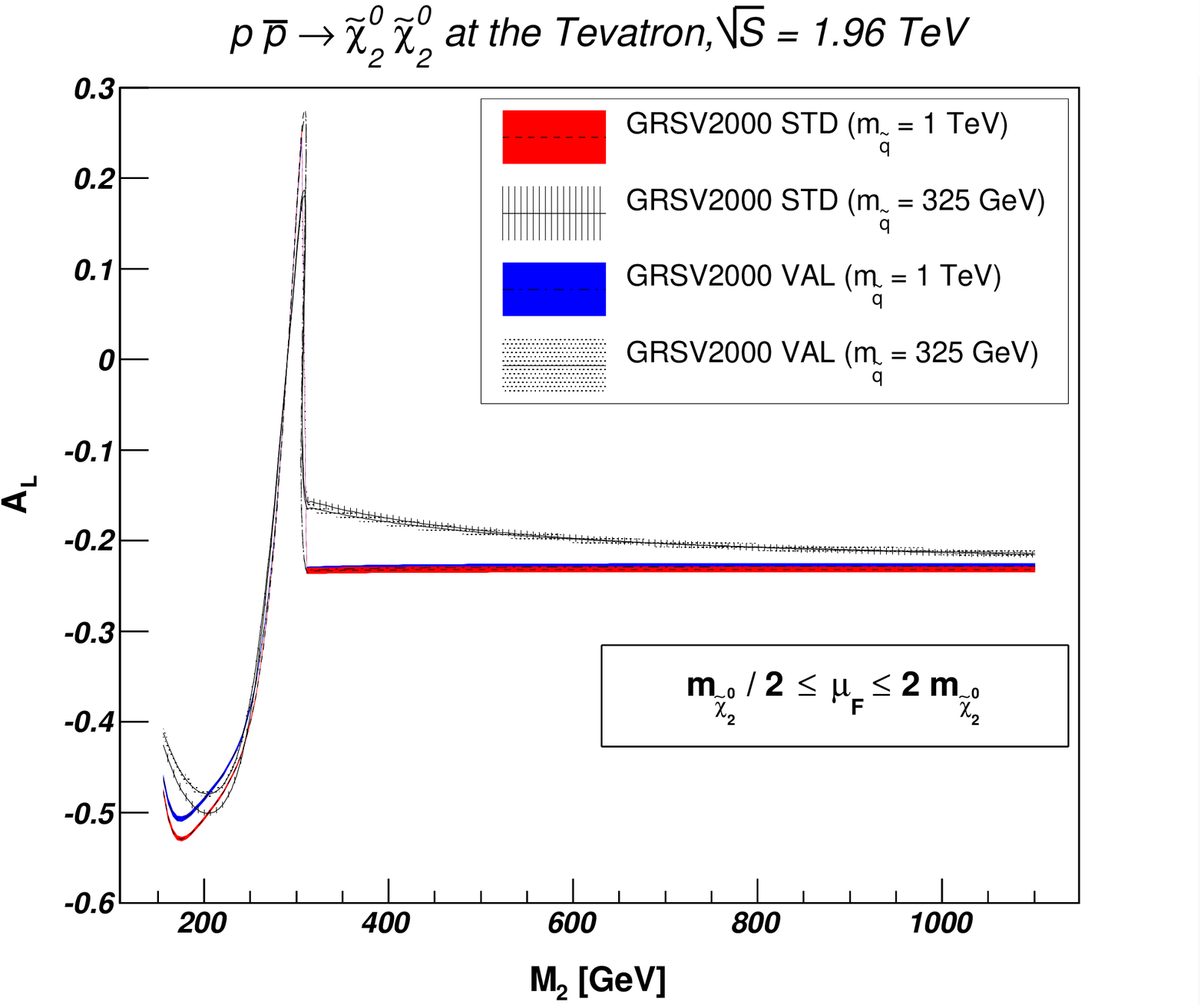,width=.49\columnwidth,clip=}
 \epsfig{file=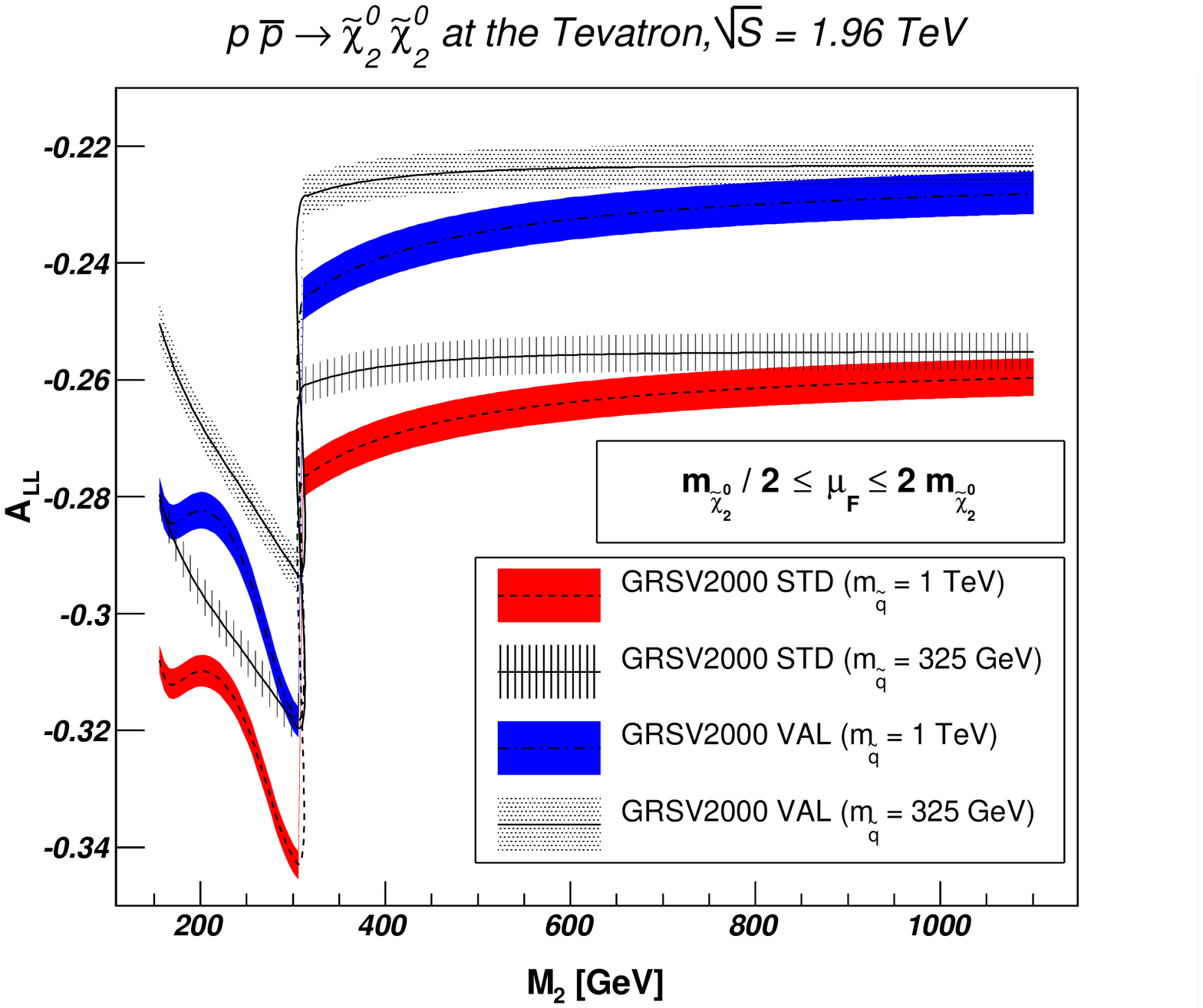,width=.49\columnwidth,clip=}
 \caption{\label{fig:6}Unpolarized gaugino-pair production cross sections
          (top left), single-spin
          asymmetries for chargino- (top right) and neutralino-pair
          production (bottom left), and double-spin asymmetries for
          neutralino-pair production (bottom right) with $m_{\tilde{\chi}
          ^0_2}\simeq m_{\tilde{\chi}^\pm_1}=151$ GeV in $p\bar{p}$
          collisions at the Tevatron and $\sqrt{S}=1.96$ TeV using LO GRV
          \cite{Gluck:1998xa} and GRSV \cite{Gluck:2000dy} parton densities.
          We choose
          $\tan\beta=10$, $\mu>0$ using Eq.\ (12), and fix $M_1={5\over3}\tan^2
          \theta_W M_2$.}
\end{figure}

In the upper left part of Fig.\ \ref{fig:6}, we show the total unpolarized
%
%
cross sections for gaugino production with $m_{\tilde{\chi}^\pm_1}=151$ GeV
at the Tevatron,
which is currently running at $\sqrt{S}=1.96$ TeV and expected to produce a
total integrated luminosity of 4--8 fb$^{-1}$ up to 2009. Therefore, besides
the pair production of the lightest chargino
(short-dashed line) and its associated production with the second-lightest
neutralino (dot-dashed line), also pair production of the latter might be
visible (long-dashed line), at least for low values of $M_2\leq300$ GeV,
where the gaugino component is still large (see Figs.\ \ref{fig:3} and
\ref{fig:4}) and the cross section exceeds
1 fb (short-dashed horizontal line). The influence of squark
exchanges and the dependence on the squark mass are
clearly visible in this channel, whereas they are again much smaller (but
slightly larger than at RHIC) for the other two channels. The factorization
scale dependence (shaded bands) remains modest (10\%-13\%) at the Tevatron.

The single-spin asymmetry for chargino-pair production (upper right part of
Fig.\ \ref{fig:6}) at a possible proton beam polarization upgrade of the
Tevatron \cite{Baiod:1995eu} could be very large and reach -40\%. Since the
physical mass has been fixed at 151 GeV and the unpolarized cross section
stays almost constant, the reduction in absolute value by about 6\% for any
given curve is directly related to the reduction of the gaugino fraction, as
$M_2$ increases. The parton density (and factorization scale) uncertainties
are (much) smaller than this variation, i.e.\ 2\% (or 1\%), so that
significant information could be extracted from this asymmetry. On the other
hand, the double-spin asymmetry (not shown), although large with about
-20\%, is almost insensitive to the gaugino fraction and would furthermore
require polarization of the antiproton beam, which is a technical challenge.

In contrast to our results for RHIC, the associated channel (not shown) is
not very interesting at the Tevatron. While the single- and double-spin
asymmetries may be large (about -10\% and +15\%, respectively), they are almost
constant and would not yield new information on the gaugino fractions.

The single- (lower left part of Fig.\ \ref{fig:6}) and double-spin
asymmetries (lower right part of Fig.\ \ref{fig:6}) for the pair production
of the second lightest neutralino are most sensitive to its gaugino
component and (relatively modest) mass variation, in particular for the low
values of $M_2\leq300$ GeV, where the cross section should be visible. Here,
$A_L$ changes sign from -50\% to almost +30\% and the theoretical
uncertainties are extremely small. In the same region, the absolute value of
$A_{LL}$ increases by about 5\% and can almost reach -35\% for the standard
GRSV parameterization of the polarized parton densities. The parton density
uncertainty remains modest with about 3\%. For large $M_2\geq300$ GeV, both
asymmetries are constant in this channel.

\subsection{LHC cross sections and asymmetries}

As the LHC is nearing completion, different upgrade scenarios are emerging,
concerning foremost higher luminosity and beam energy \cite{Zimmermann:2007zz},
but also beam polarization \cite{roeck}. It is interesting to remember that a
detailed study has been performed some time ago for the SSC, resulting in a
design that had reserved 52 lattice locations for the future installation
of Siberian snakes \cite{Krisch:1998zm}. Since this is currently not the case
at the LHC, its polarization upgrade would require replacing some of the
dipoles with higher-field magnets to create these locations, just as in the case
of the Tevatron. The number of resonances to be crossed during acceleration
would be considerably larger due to the higher energy of the LHC, requiring
longer tuning before ultimately reaching polarizations of up to 65\%-70\%.

\begin{figure}
 \centering
 \epsfig{file=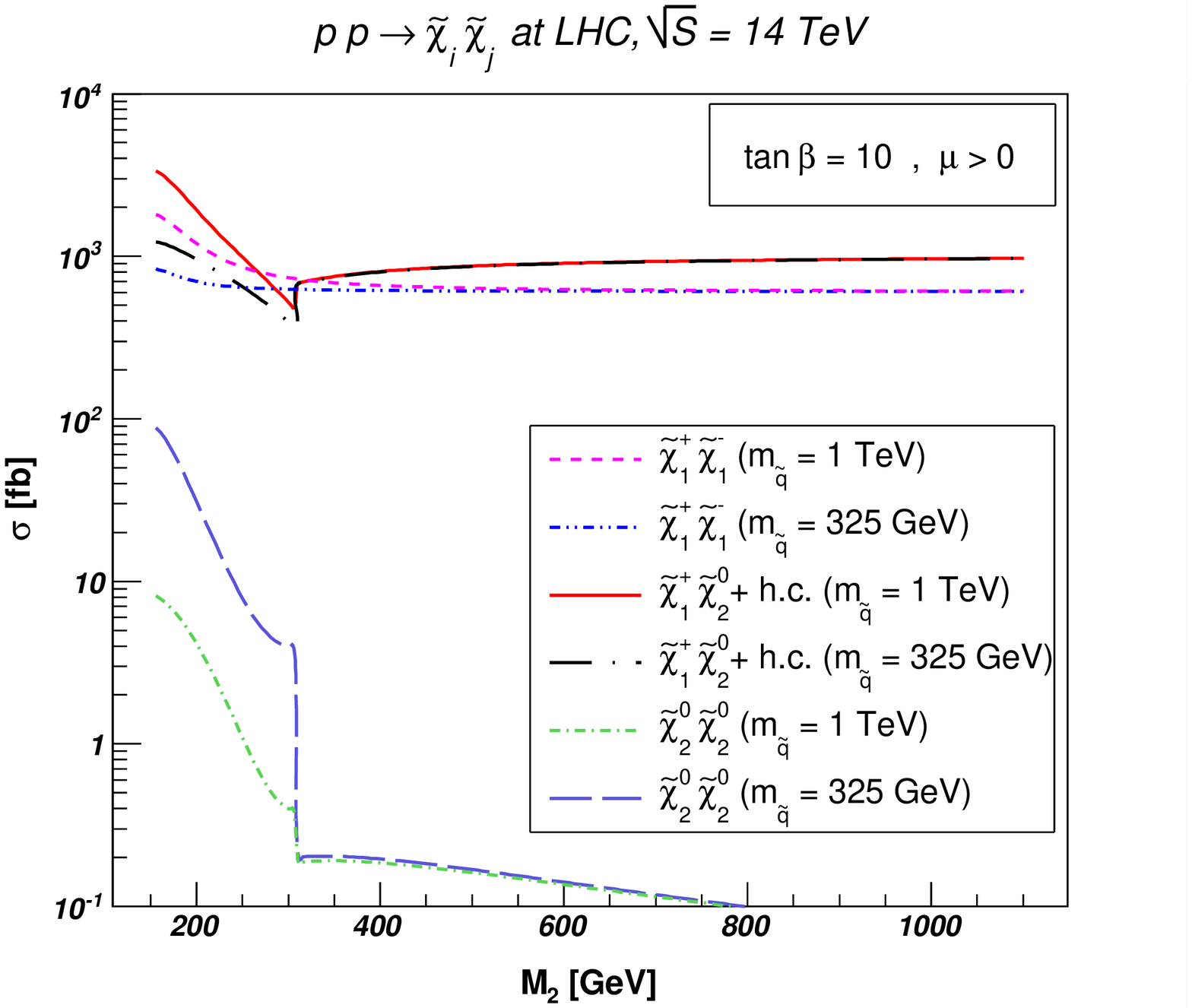,width=.49\columnwidth,clip=}
 \epsfig{file=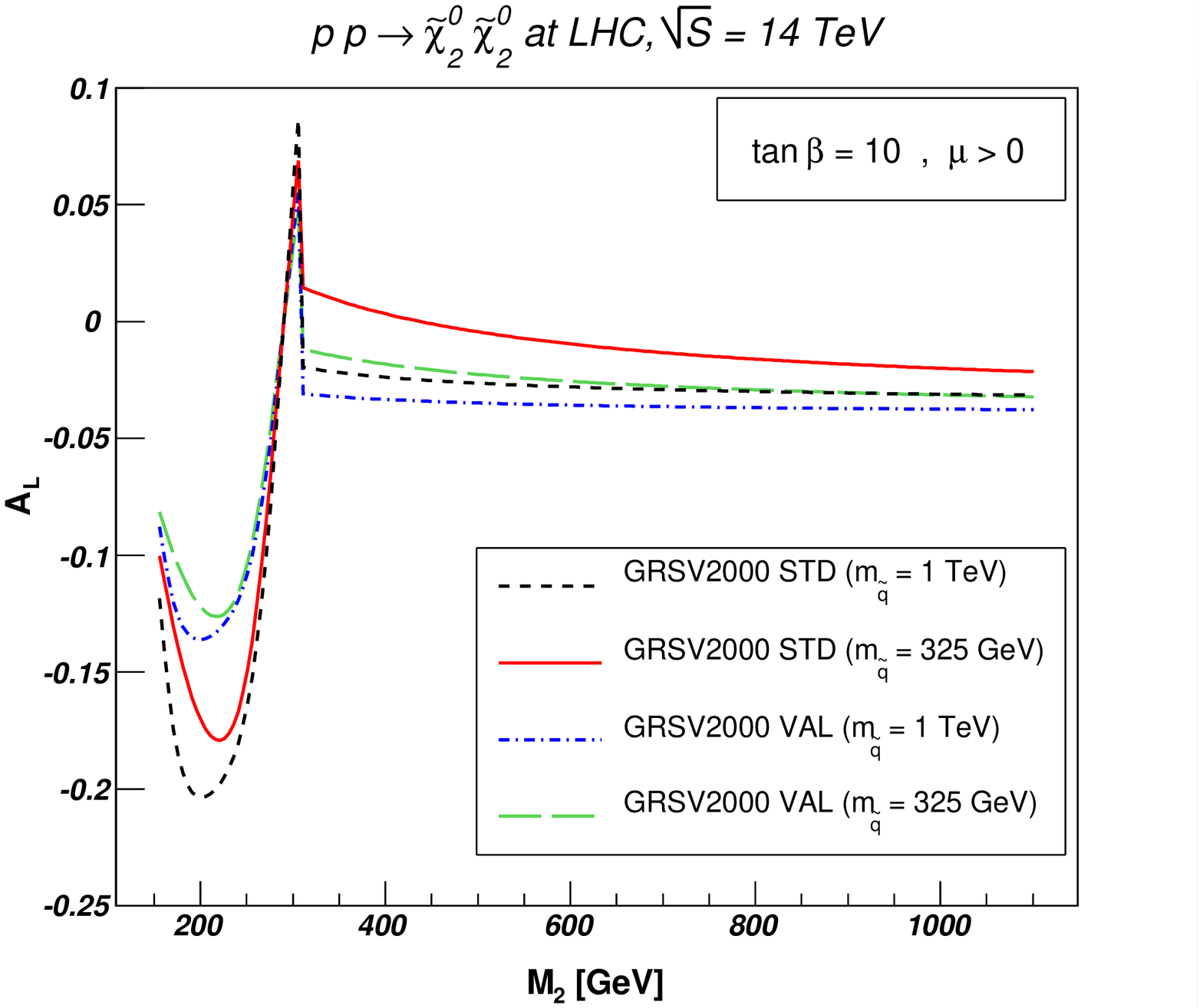,width=.49\columnwidth,clip=}
 \epsfig{file=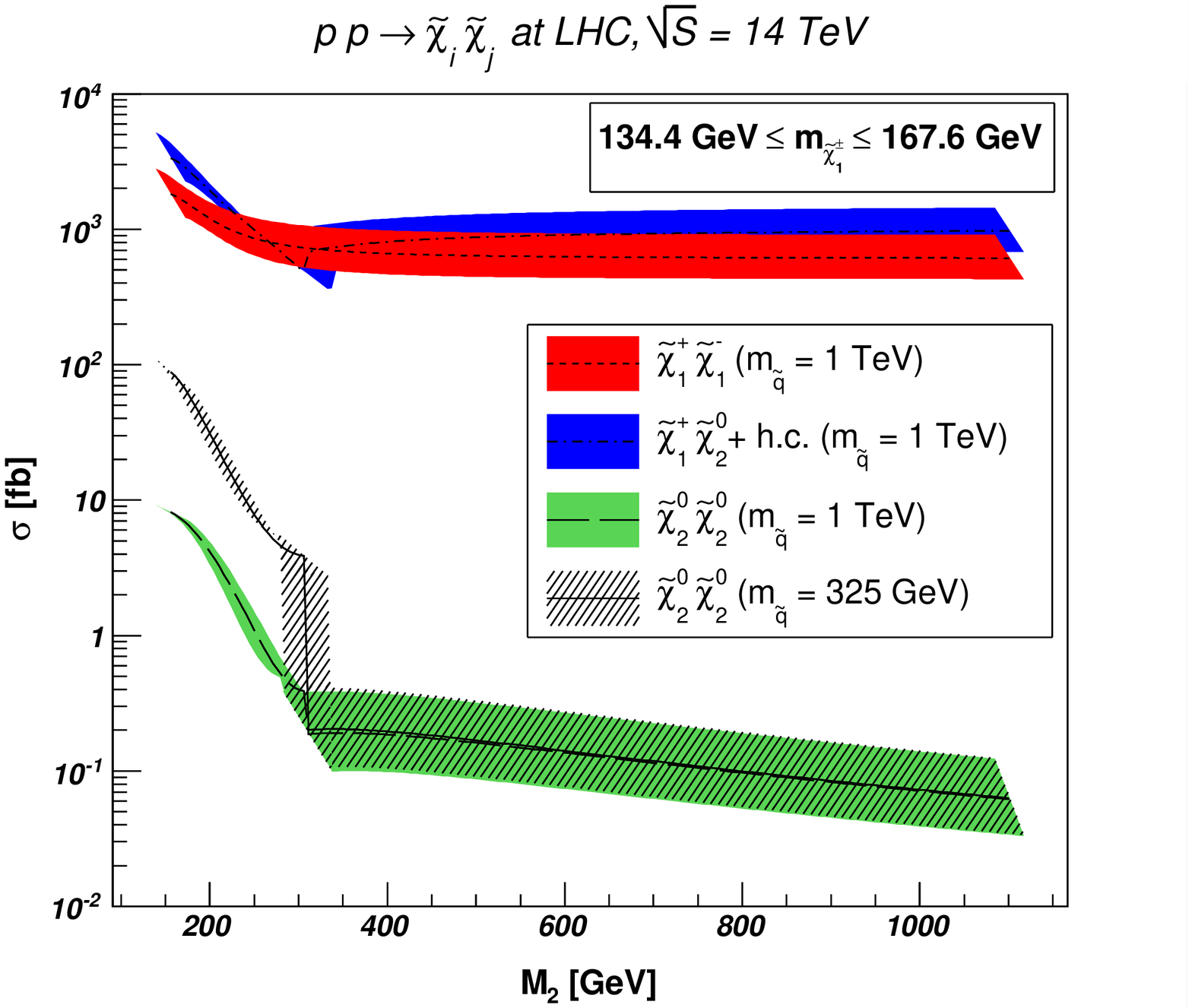,width=.49\columnwidth,clip=}
 \epsfig{file=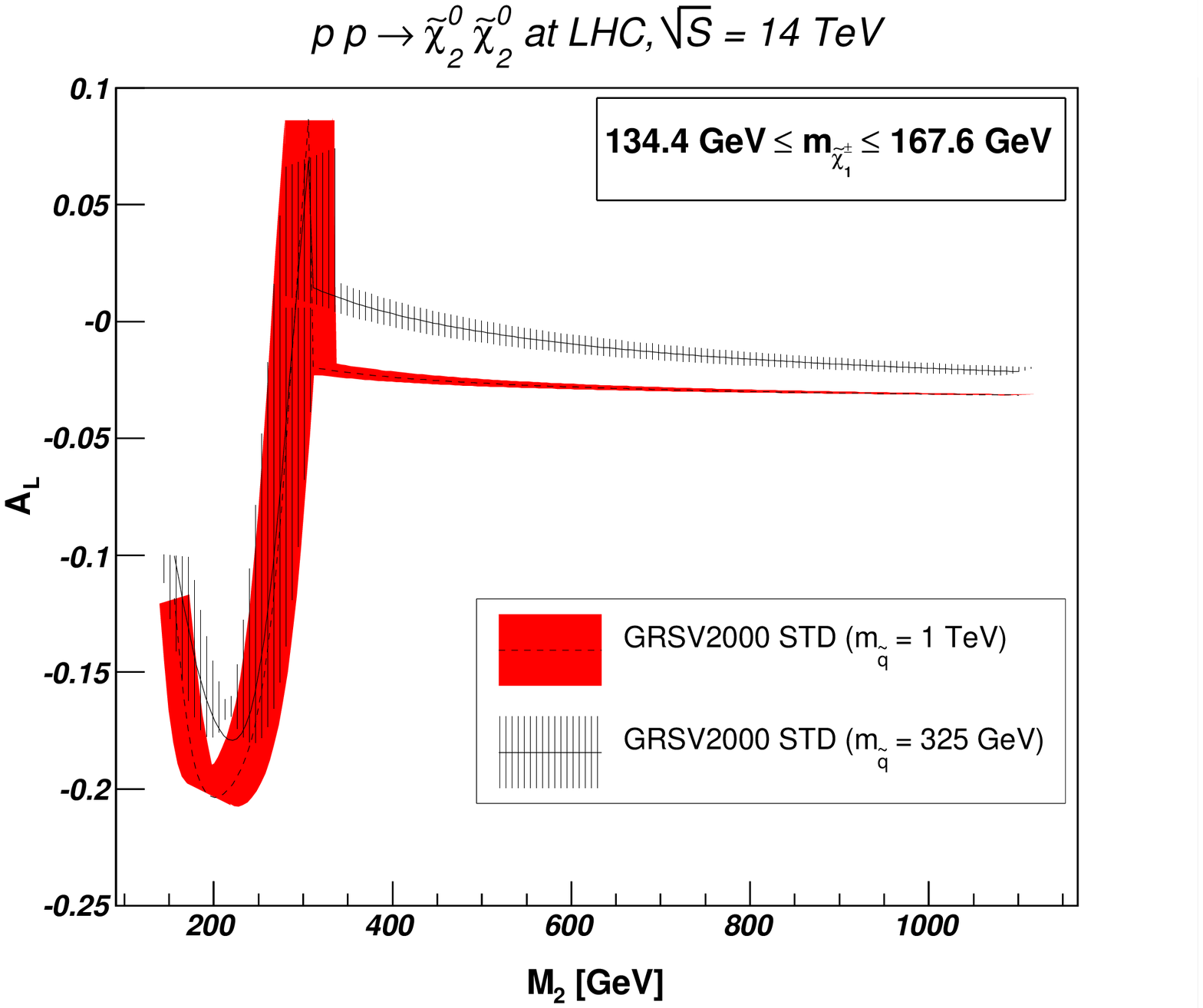,width=.49\columnwidth,clip=}
 \caption{\label{fig:7}Unpolarized gaugino-pair production cross sections
          (left) and single-spin
          asymmetries for neutralino-pair production (right) with
          $m_{\tilde{\chi}^0_2}\simeq m_{\tilde{\chi}^\pm_1}=151$ GeV in
          $pp$ collisions at the LHC and $\sqrt{S}=14$ TeV using LO GRV
          \cite{Gluck:1998xa} and GRSV \cite{Gluck:2000dy} parton densities.
          The shaded bands
          (bottom) show the uncertainty induced by the error on the chargino
          mass as determined with 100 fb$^{-1}$ of data \cite{Nojiri:2003tv}.
          We choose
          $\tan\beta=10$, $\mu>0$ using Eq.\ (12), and fix $M_1={5\over3}\tan^2
          \theta_W M_2$.}
\end{figure}

For $pp$ collisions of 14 TeV center-of-mass energy at the LHC, we show the
unpolarized total cross sections for a chargino of mass 151 GeV in the upper left
part of Fig.\ \ref{fig:7}. Low-luminosity LHC runs of 10 fb$^{-1}$ per year
%
%
are currently scheduled for the years 2009 to 2011, and the high-luminosity
phase of 100 fb$^{-1}$ per year
should start in 2011. With these luminosities, pair
production of the lightest chargino (short-dashed line), its associated
production with the second-lightest neutralino (dot-dashed line), and pair
production of the latter (long-dashed line) should all be well visible.
Whereas the cross sections for the first two channels are again almost
constant and fairly independent of the squark mass, at least for the
Higgsino-like region of $M_2\geq300$ GeV, the neutralino-pair production
cross section is again quite sensitive to squark exchanges in the
gauginolike region below that value and stays almost constant above. The
factorization scale dependence is very small at the LHC and included in
the line width of the upper left part of Fig.\ \ref{fig:7}. Note, however,
that with 100 fb$^{-1}$ of data, the mass of the lightest chargino will
only be measured with an uncertainty of $\pm11$\% \cite{Nojiri:2003tv}.
This induces a very visible uncertainty (shaded bands) in the total cross
sections (lower left part of Fig.\ \ref{fig:7}).

For a possible polarization upgrade of the LHC \cite{roeck}, we show the
single-spin asymmetry for neutralino-pair production in the right parts of
Fig.\ \ref{fig:7}, again with the scale (line width, top) and chargino mass
(shaded bands, bottom) uncertainty.
At $M_2\geq300$ GeV, where the gaugino fraction is small,
the asymmetry is not very interesting, as it is almost constant and smaller
than 5\%. In the gauginolike region at $M_2\leq300$ GeV, it changes sign
from -20\% to almost +10\%,
a variation, that is considerably larger than the parton density uncertainty
of at most 7\%, the squark mass dependence of at most 2\%, the almost
invisible scale dependence, and also the chargino mass uncertainty of 3\% to
10\%. At a polarized LHC, a measurement of the
single-spin asymmetry for neutralino-pair production would therefore yield
interesting information about its gaugino fraction.

\begin{figure}
 \centering
 \epsfig{file=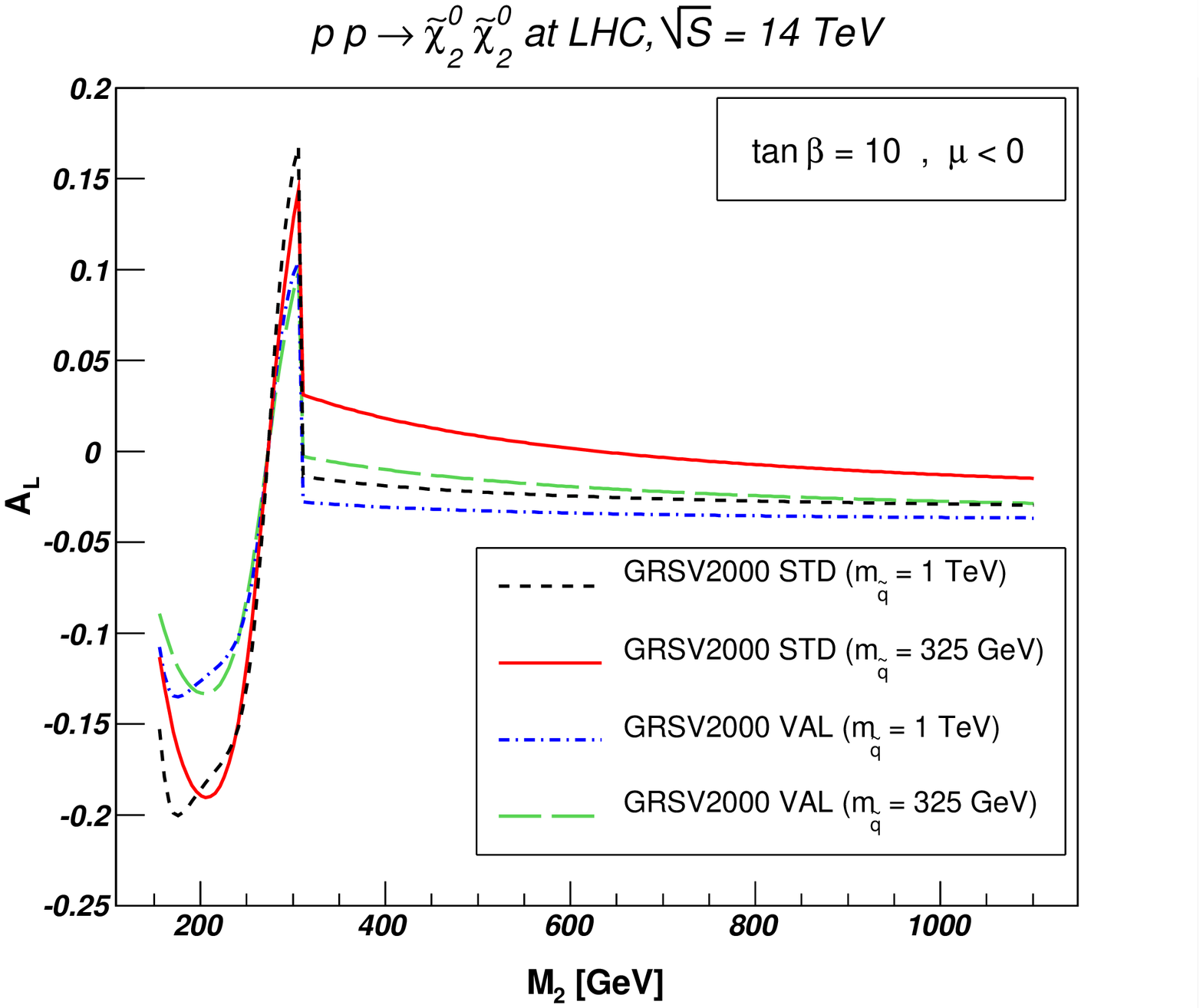,width=.49\columnwidth,clip=}
 \epsfig{file=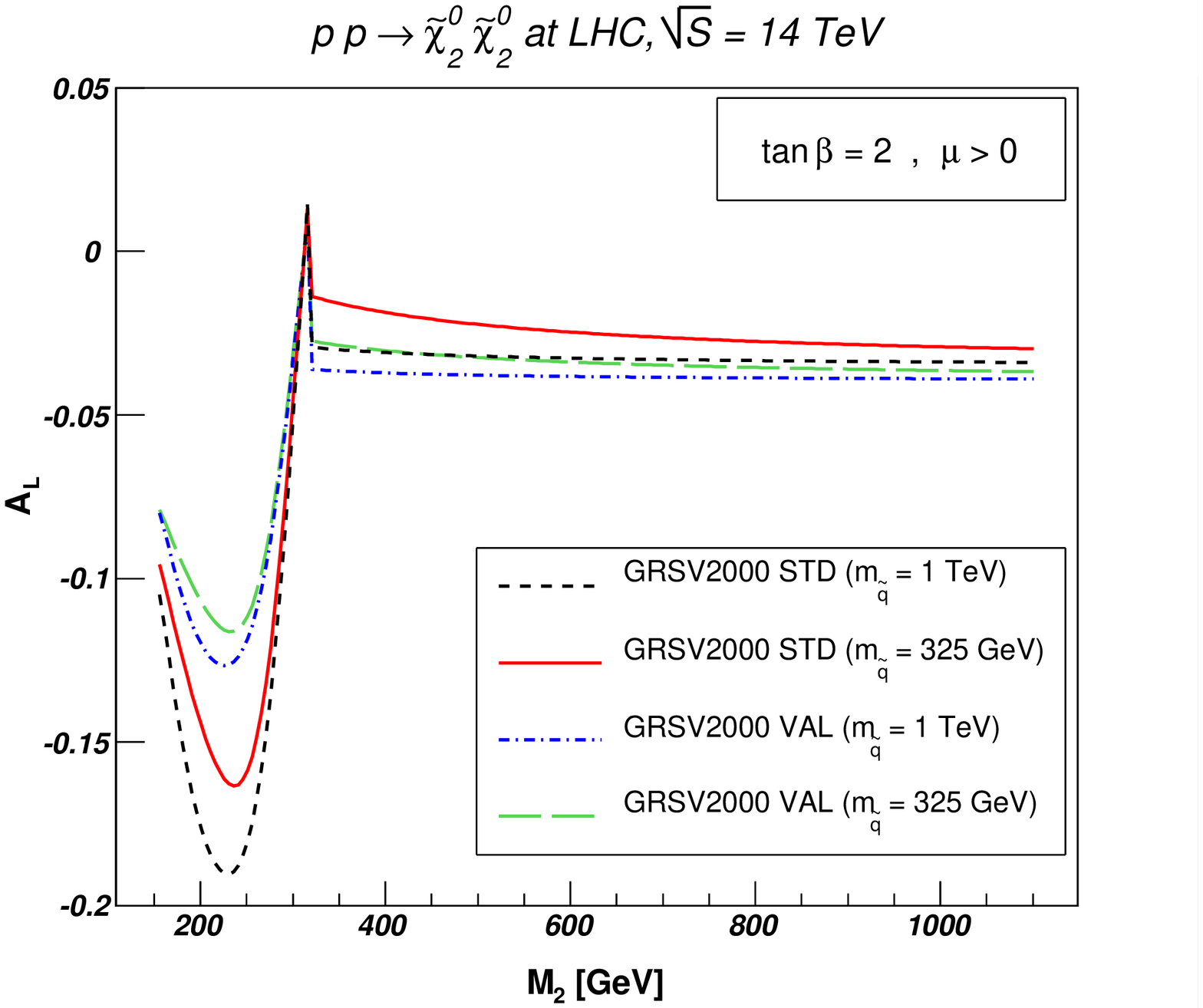,width=.49\columnwidth,clip=}
 \epsfig{file=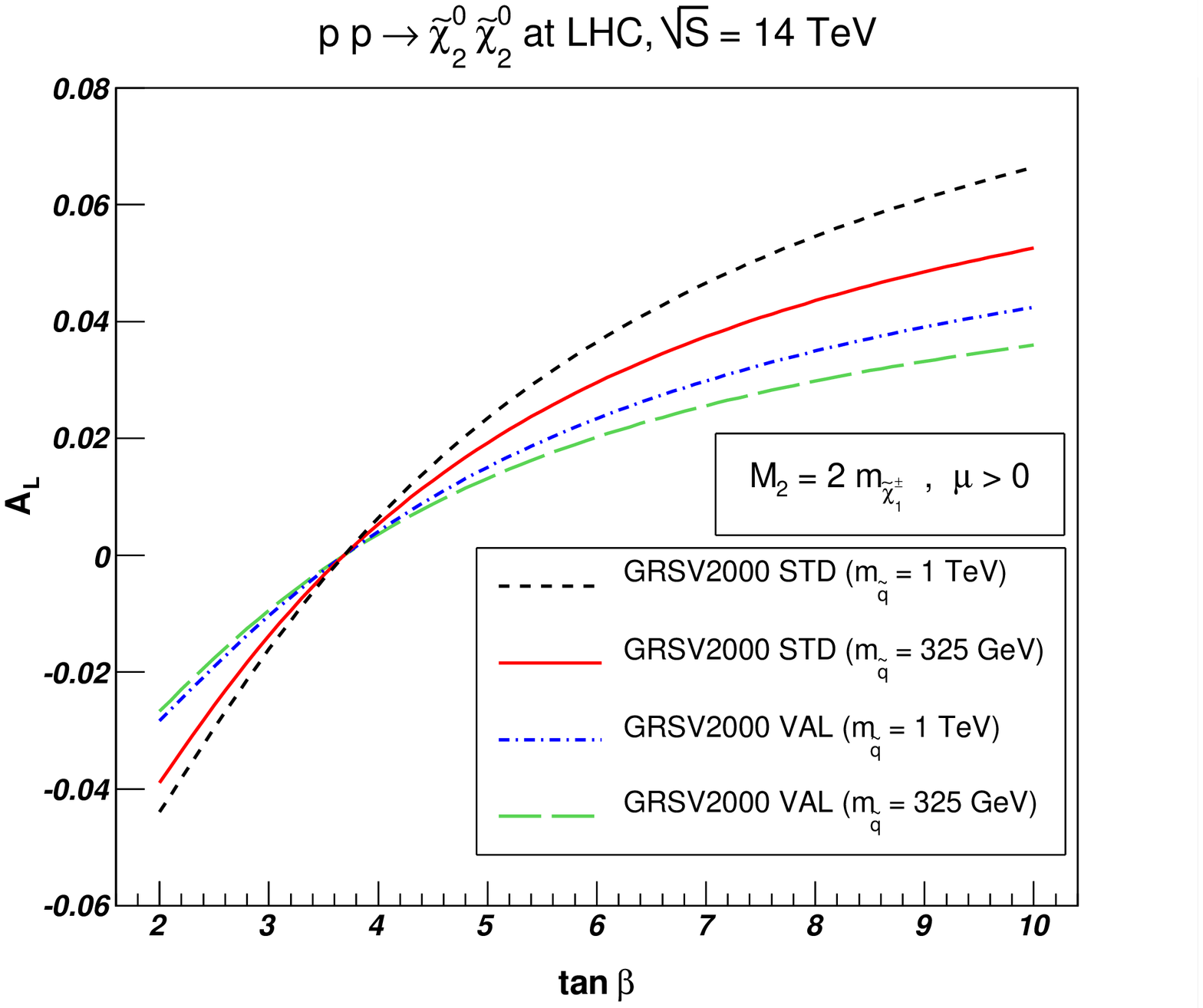,width=.49\columnwidth,clip=}
 \epsfig{file=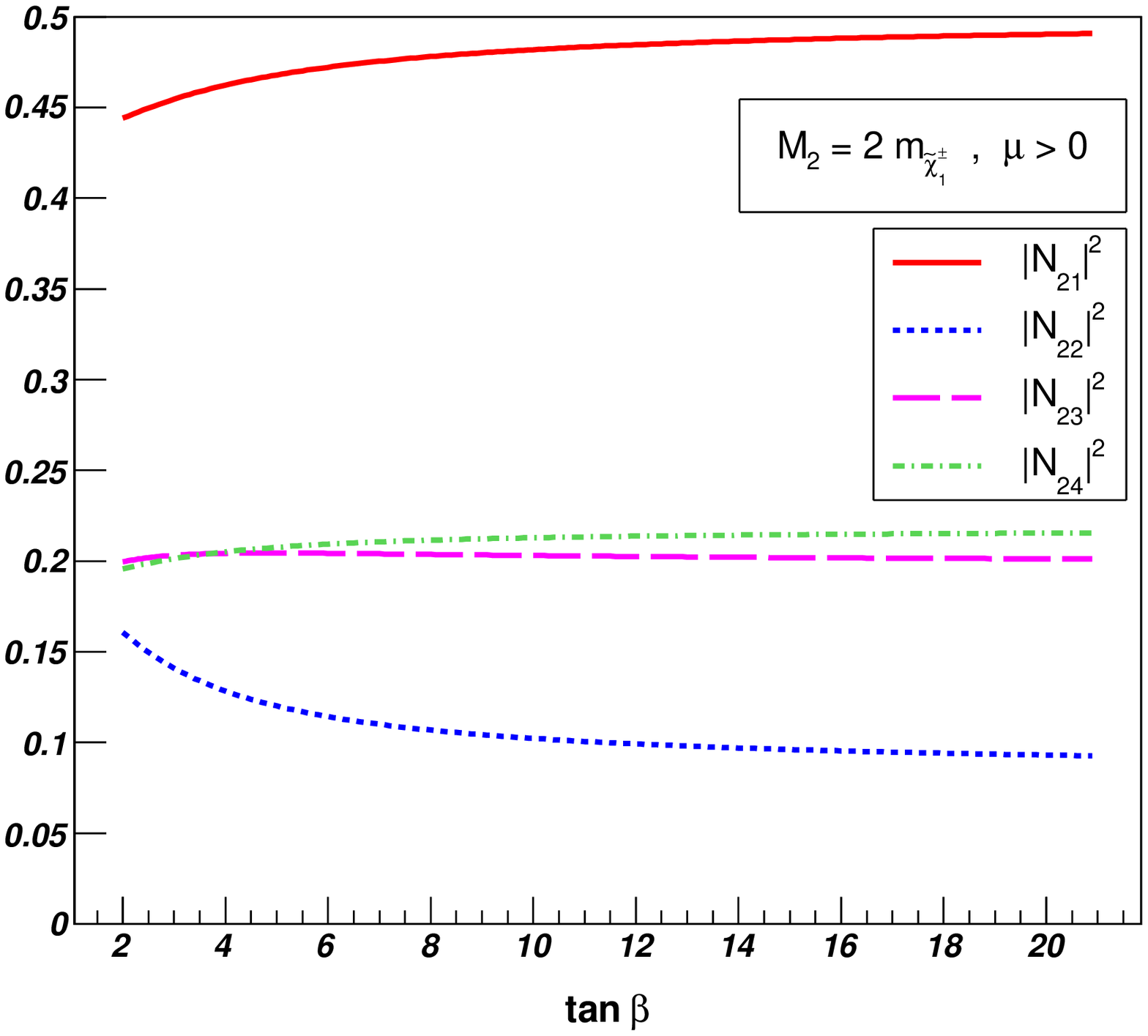,width=.49\columnwidth,clip=}
 \caption{\label{fig:8}Single-spin asymmetries for neutralino-pair production
          with $\tan\beta=10$ and $\mu<0$ (top left), $\tan\beta=2$ and $\mu>0$
          (top right), and $M_2=2 m_{\tilde{\chi}^\pm_1}$ with $\mu>0$ as a
          function of $\tan\beta$ (bottom left) in $pp$ collisions at the LHC and
          $\sqrt{S}=14$ TeV. For the third scenario, we show also the gaugino
          and Higgsino fractions of the second-lightest neutralino (bottom right).
          We fix $\mu$ using Eq.\ (12) and $M_1={5\over3}\tan^2 \theta_W M_2$.}
\end{figure}

While the cross sections vary very little when changing the sign of $\mu$
or varying $\tan\beta$, it is interesting to study further the single-spin
asymmetries for neutralino pairs in these alternative scenarios. When comparing
the asymmetry for $\mu<0$, shown in the upper left part of Fig.\ \ref{fig:8},
%
%
to the one for $\mu>0$, shown in the upper right part of Fig.\ \ref{fig:7},
one notices an even steeper rise in the former to more than +15\%, as $M_2$
approaches the critical value of 2$m_{\tilde{\chi}_1^\pm}$, the neutralino
changes its character from gaugino to Higgsino, and the sign and (smaller)
absolute value of the Higgsino mass parameter $\mu$ become of particular
importance.

A similar effect is observed when comparing for $\mu>0$ the asymmetry with
a lower value of $\tan\beta=2$ in the upper right part of Fig.\ \ref{fig:8} to
the one for the standard value of $\tan\beta=10$ in the upper right part of
Fig.\ \ref{fig:7}. In this case, the asymmetry rises less (barely above zero)
towards $M_2=2m_{\tilde{\chi}_1^\pm}$, where the gaugino/Higgsino decomposition
is flipped, the ratio of the two Higgs vacuum expectation values $\tan\beta$ is
particularly important, and the absolute value of the Higgsino mass parameter
$\mu$ is effectively larger than in the standard scenario.

The dependence on $\tan\beta$ at the critical point $M_2=2m_{\tilde{\chi}_1^+}$
can be seen more clearly in the lower left part of Fig.\ \ref{fig:8}, and
indeed the asymmetry decreases from large to small $\tan\beta$ from
distinctively positive values to values at or below zero for all choices of
squark masses (1 TeV or 325 GeV) and parton density functions (standard or
valence GRSV parameterizations). This decrease is correlated with a similar
decrease in the $B$-ino fraction $|N_{21}|^2$ and with an increase in the
$W$-ino fraction $|N_{22}|^2$ of the second-lightest neutralino, while the
Higgsino fractions $|N_{23}|^2$ and $|N_{24}|^2$ stay almost constant, as
can be seen in the lower right part of Fig.\ \ref{fig:8}.

The double-spin asymmetry for neutralino pairs, as well as the one for
chargino pairs and the associated channel, are always smaller than 4\% and
2\%, respectively. Furthermore, they vary by less than 2\% and are therefore
not shown here. The single-spin asymmetry for chargino pairs (not shown) can
reach a slightly larger value of -12\%, but again it varies by less than 3\%
as a function of $M_2$, which is almost of the same size as the parton
density uncertainty (2\%). The situation for the single-spin asymmetry of
the associated channel (not shown, either) is similar with a maximum of
-10\%, a variation with $M_2$ of about 1\% and a parton density uncertainty
of less than 1\%.

\section{Conclusion}
\label{sec:4}

In summary, we have presented an exploratory study of gaugino-pair
production in polarized hadron collisions at RHIC and possible upgrades of
the Tevatron and the LHC, focusing on the correlation of beam polarization
and gaugino/Higgsino mixing in the MSSM. Assuming gaugino mass unification
at the GUT scale, a typical value of $\tan\beta=10$ (or 2), a known lightest
chargino mass and restricting ourselves (mostly) to positive values of $\mu$,
favored by the anomalous magnetic moment of the muon, we computed gaugino
cross sections and beam polarization asymmetries as a function of the
SUSY-breaking mass parameter $M_2$ without imposing a particular
SUSY-breaking model.

While the unpolarized cross sections were held almost constant by imposing a
fixed chargino mass, the single-spin asymmetries were found to be strongly
correlated with $M_2$ and therefore the gaugino fractions of the lightest
chargino and second-lightest neutralino in their associated production at
RHIC, where the asymmetries could reach -20\%. Even larger asymmetries of up
to -40\% would be obtained at a polarization upgrade of at least the proton
beam of the Tevatron, where the single-spin asymmetry of chargino-pair
production and the single- and double-spin asymmetries of neutralino-pair
production would be the most promising observables. At the LHC, proton beam
polarization would make it possible to measure in particular the single-spin
asymmetry of neutralino pairs, which changes sign as $M_2$ grows and the
gaugino fraction of the second-lightest neutralino falls.

For the channels mentioned above, the theoretical uncertainties coming from
parton density, factorization scale, and squark mass variations and a realistic
experimental uncertainty on the lightest chargino mass were found
to be smaller than the differences induced in the asymmetries by variations
of the gaugino fractions. However, more information on the so far poorly
constrained polarized parton densities would clearly increase the impact
of an analysis of gaugino-pair production in polarized hadron collisions.

\acknowledgments
We thank A.\ de Roeck, M.\ Drees, S.\ Kraml, P.\ Newman, T.\ Roser, I.\
Schienbein and W.\ Vogelsang for useful discussions.
This work has been supported by a Ph.D.\ fellowship of the French ministry
for education and research and by the Theory-LHC-France initiative of the
CNRS/IN2P3.

\appendix

\section{Gaugino and Higgsino mixing}
\label{sec:a}

The soft SUSY-breaking terms in the minimally supersymmetric Lagrangian
include a term \cite{Haber:1984rc}
\bea
 {\cal L}&\supset&-{1\over2}(\psi^0)^T\,Y\,\psi^0+{\rm h.c.},
\eea
which is bilinear in the (two-component) fermionic partners
\bea
 \psi^0_j~=~(-i\tilde{B},-i\tilde{W}^3,\tilde{H}_1^0,\tilde{H}_2^0)^T
 ~~ &{\rm with}& ~~ j=1,\dots,4
\eea
of the neutral electroweak gauge and Higgs bosons and proportional to the,
generally complex and necessarily symmetric, neutralino mass matrix
\bea
 Y &=& \left( \begin{array}{cccc}
  M_1 & 0 &
  -m_Z\,s_W\,c_\beta &
  ~~m_Z\,s_W\,s_\beta \\
  0 & M_2 &
  ~~m_Z\,c_W\,c_\beta &
  -m_Z\,c_W\,s_\beta \\
  -m_Z\,s_W\,c_\beta &
  ~~m_Z\,c_W\,c_\beta &
  0 & -\mu \\
  ~~m_Z\,s_W\,s_\beta &
  -m_Z\,c_W\,s_\beta &
  -\mu & 0
 \end{array} \right).
\eea
Here, $M_1$, $M_2$, and $\mu$ are the SUSY-breaking $B$-ino, $W$-ino, and
off-diagonal Higgsino mass parameters with $\tan\beta=s_\beta/c_\beta=v_u/
v_d$ being the ratio of the vacuum expectation values $v_{u,d}$ of the two
Higgs doublets, while $m_Z$ is the SM $Z$-boson mass and $s_W$ $(c_W)$ is
the sine (cosine) of the electroweak mixing angle $\theta_W$. After
electroweak gauge-symmetry breaking and diagonalization of the mass matrix
$Y$, one obtains the neutralino mass eigenstates
\bea
 \chi^0_i&=&N_{ij}\,\psi_j^0,~~~i=1,\dots,4,
\eea
where $N$ is a unitary matrix satisfying the relation
\bea
 N^*\,Y\,N^{-1}&=&{\rm diag}\, (m_{\tilde{\chi}^0_1},m_{\tilde{\chi}^0_2},
 m_{\tilde{\chi}^0_3},m_{\tilde{\chi}^0_4}).
\eea
In four-component notation, the Majorana-fermionic neutralino mass eigenstates
can be written as
\bea
 \tilde{\chi}^0_i&=&\lr\begin{array}{c} \chi_i^0\\
 \bar{\chi}_i^0\end{array}\rr.
\eea
The application of projection operators leads to relatively compact analytic
expressions for the mass eigenvalues $m_{\tilde{\chi}^0_1}<
m_{\tilde{\chi}^0_2}<m_{\tilde{\chi}^0_3}<m_{\tilde{\chi}^0_4}$
\cite{Gounaris:2001fx}. As we choose them to be real and non-negative, our
unitary matrix $N$ is generally complex \cite{ElKheishen:1992yv}.
For the MSSM with additional $CP$-violating phases see Ref.\ \cite{Ibrahim:2007fb}
and the references therein.

The chargino mass term in the SUSY Lagrangian \cite{Haber:1984rc}
\bea
 {\cal L} &\supset& -{1\over2} (\psi^+\psi^-)\lr\begin{array}{cc}
 0&X^T\\X&0\end{array}\rr \lr\begin{array}{c}\psi^+\\ \psi^-\end{array}\rr
 +{\rm h.c.}
\eea
is bilinear in the (two-component) fermionic partners
\bea
 \psi_j^\pm~=~(-i\tilde{W}^\pm,\tilde{H}^\pm_{2,1})^T
 ~~ &{\rm with}& ~~ j=1,\dots,2
\eea
of the charged electroweak gauge and Higgs bosons and proportional to the,
generally complex, chargino mass matrix
\bea
 X &=& \left( \begin{array}{c c} M_{2} & m_{W}\, \sqrt{2}\, s_\beta \\
 m_{W}\, \sqrt{2}\, c_\beta &  \mu \end{array}\right),
\eea
where $m_W$ is the mass of the SM $W$-boson. Since $X$ is not symmetric, it
must be diagonalized by two unitary matrices $U$ and $V$, which satisfy the
relation
\bea
 U^*\,X\,V^{-1}&=&{\rm diag}\,(m_{\tilde{\chi}^\pm_1},m_{\tilde{\chi}^
 \pm_2})\label{eq:a10}
\eea
and define the chargino mass eigenstates
\bea
 \begin{array}{l} \chi_i^+~=~V_{ij}\,\psi_j^+\\
 \chi_j^-~=~U_{ij}\,\psi_j^-\end{array},~~~i,j=1,2.
\eea
In four-component notation, the Dirac-fermionic chargino mass eigenstates can
be written as
\bea
 \tilde{\chi}^\pm_i&=&\lr\begin{array}{c} \chi_i^\pm\\
 \bar{\chi}_i^\mp\end{array}\rr.
\eea
As Eq.\ (\ref{eq:a10}) implies
\bea
 VX^\dagger XV^{-1}&=&{\rm diag}\,(m_{\tilde{\chi}^\pm_1}^2,m_{\tilde{\chi}^
 \pm_2}^2),
\eea
the hermitian matrix $X^\dagger X$ can be diagonalized using only $V$, and
its eigenvalues
\bea
 m_{\tilde{\chi}^\pm_{1,2}}^2 &=& \frac{1}{2}\left\{|M_2|^2+|\mu|^2+
 2 m_W^2 \mp \sqrt{(|M_2|^2+|\mu|^2+ 2 m_W^2)^2 - 4 |\mu M_2- m_W^2
 s_{2\beta}|^2}\right\}
 \label{eq:a14}
\eea
are always real. If we take also the mass eigenvalues $m_{\tilde{\chi}^\pm_
1}\leq m_{\tilde{\chi}^\pm_2}$ to be real and non-negative, the rotation
matrix
\bea
 V&=&\lr\begin{array}{ll}~~\,\cos\theta_+&\sin\theta_+\,e^{-i\phi_+}\\
                        -\sin\theta_+\,e^{i\phi_+}&\cos\theta_+\end{array}\rr
\eea
can still be chosen to have real diagonal elements, but the off-diagonal
phase $e^{\mp i\phi_+}$ is needed to rotate away the imaginary part of the
off-diagonal matrix element in $X^\dagger X$,
\bea
 \Im \le (M_2^*\,s_\beta+\mu\,c_\beta)\,e^{i\phi_+}\re &=& 0.
\eea
The rotation angle $\theta_+\in[0;\pi]$ is uniquely fixed by the two
conditions
\bea
 \tan2\theta_+&=&\frac{2\sqrt{2}m_W \left( M_2^*\,s_\beta + \mu\,c_\beta
 \right)\,e^{i\phi_+}} {|M_2|^2 -|\mu|^2 + 2 m_W^2c_{2\beta}}~~~~{\rm and}\\
 \sin2\theta_+&=&\frac{-2\sqrt{2} m_W \left( M_2^*\,s_\beta + \mu\,c_\beta
 \right)\,e^{i\phi_+}
 }
 {\sqrt{(|M_2|^2 -|\mu|^2 + 2 m_W^2c_{2\beta})^2
 +8 m_W^2\le(M_2^*\, s_\beta + \mu\,c_\beta)\,e^{i\phi_+}\re^2}}.
\eea
Once $V$ is known, the unitary matrix $U$ can be obtained from
\bea
 U &=& {\rm diag}\,(m_{\tilde{\chi}^\pm_1}^{-1},m_{\tilde{\chi}^\pm_2}^{-1})
 ~ V^*X^T.
\eea
For the MSSM with additional $CP$-violating phases see again Ref.\
\cite{Ibrahim:2007fb} and the references therein.

\section{Feynman rules}
\label{sec:b}

\begin{figure}
 \centering
 \epsfig{file=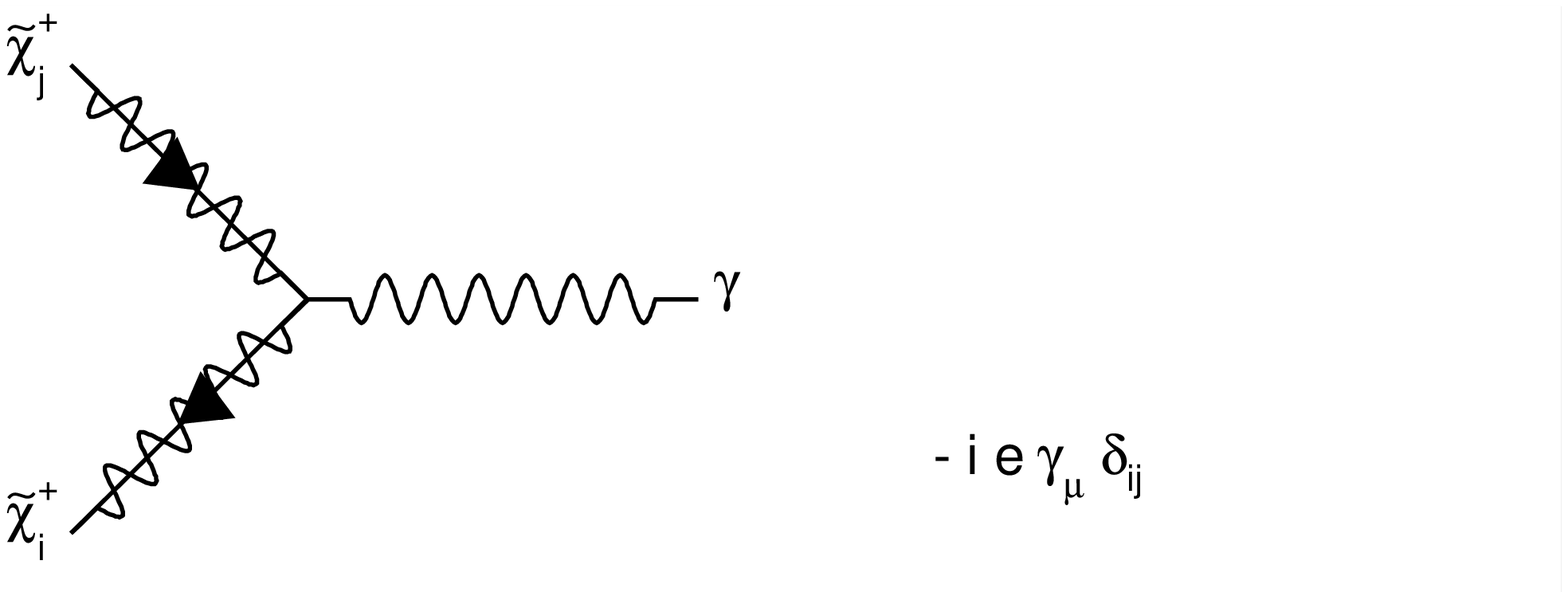,width=.49\columnwidth}
 \epsfig{file=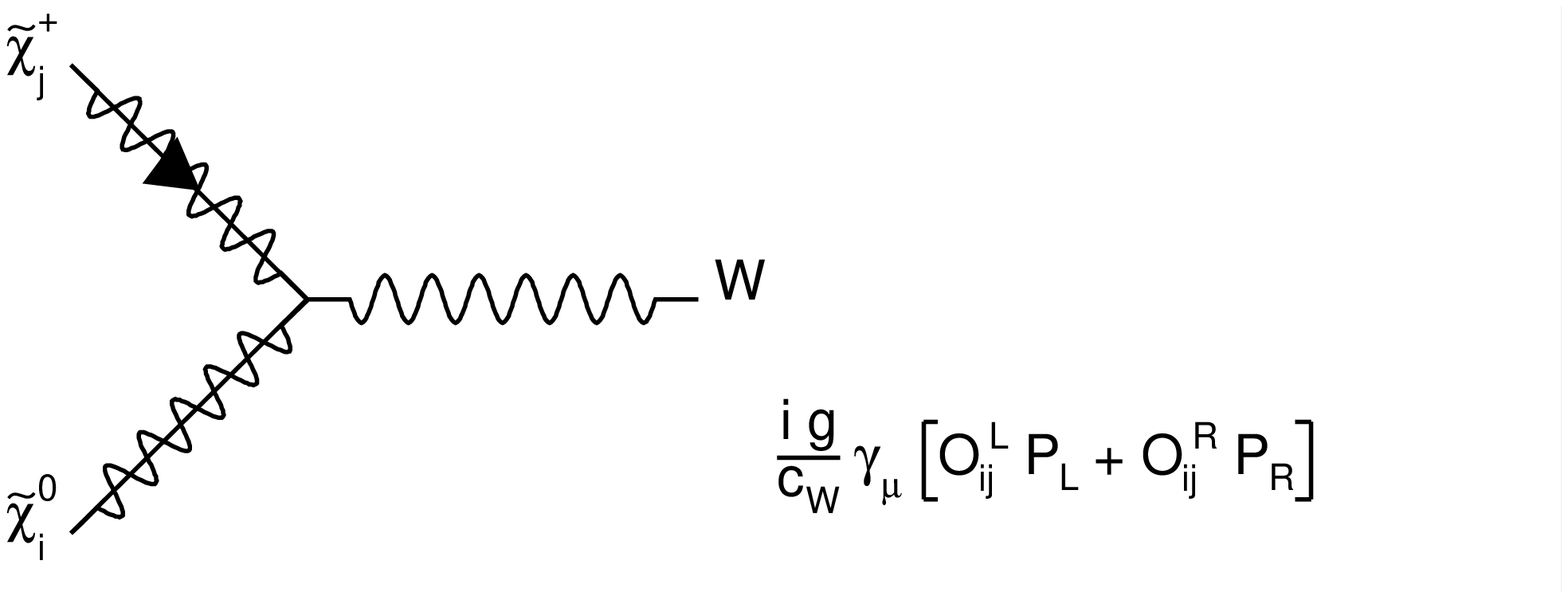,width=.49\columnwidth}
 \epsfig{file=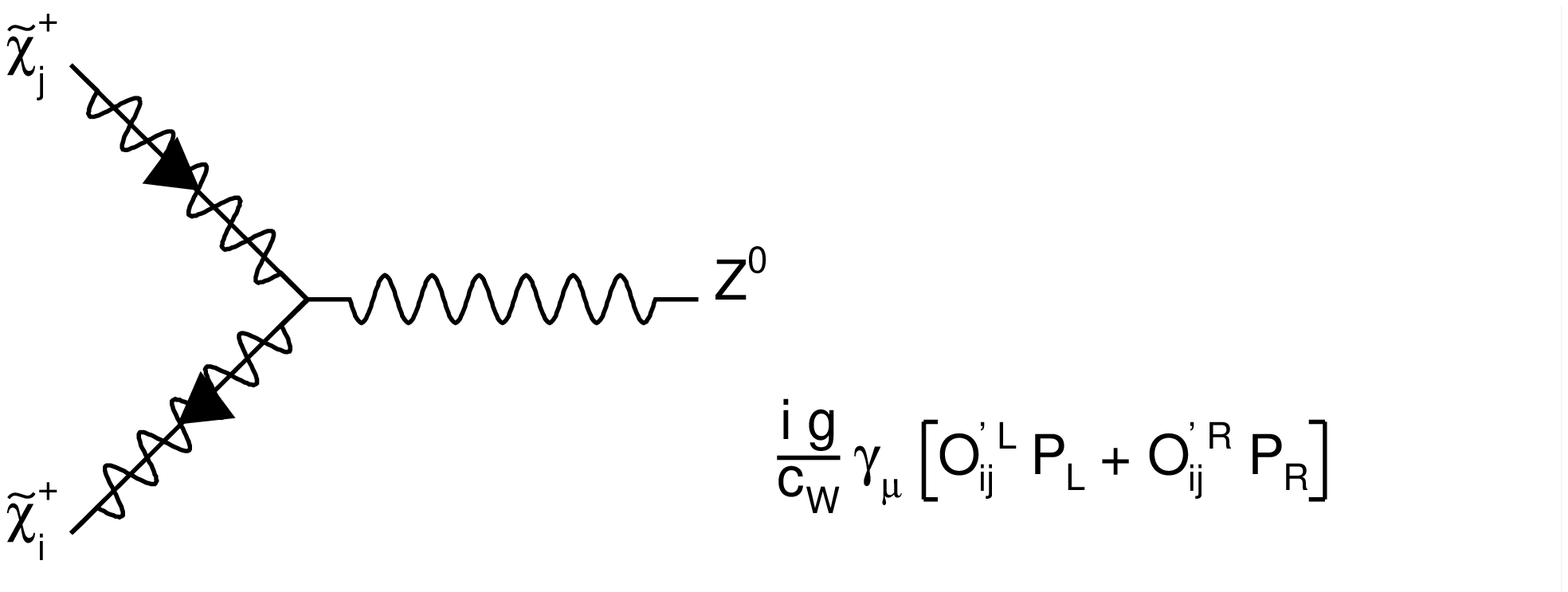,width=.49\columnwidth}
 \epsfig{file=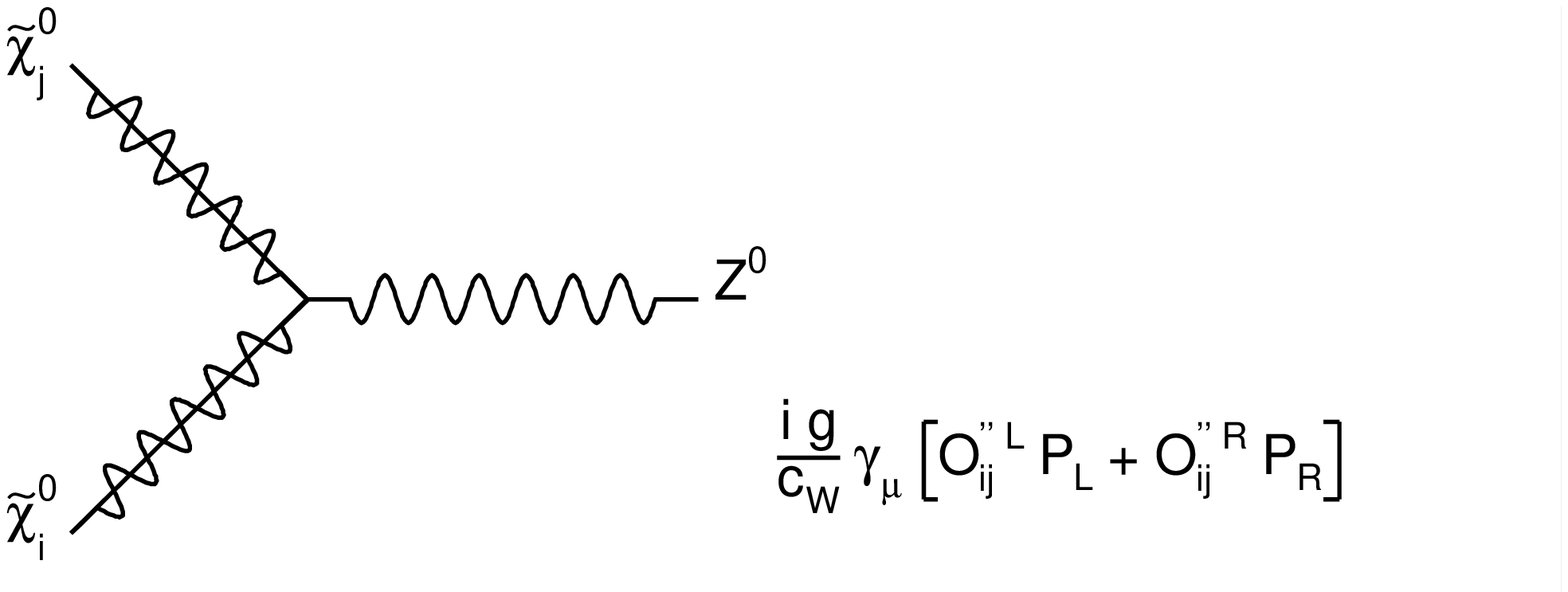,width=.49\columnwidth}
 \caption{\label{fig:b1}Feynman rules for interactions of charginos and
 neutralinos with electroweak gauge bosons. The arrows denote the direction
 of electric charge flow: $+1$ in the case of $\tilde{\chi}^+$ and $e_q$ 
 in the case of $q$ and $\tilde{q}$ (see below).}
\end{figure}

For the electroweak interaction, we define the square of the weak coupling
constant $g^2=e^2/\sin^2\theta_W$ in terms of the electromagnetic fine
structure constant $\alpha=e^2/(4\pi)$ and the squared sine of the
electroweak mixing angle $x_W=\sin^2\theta_W=s_W^2 = 1-\cos^2\theta_W =
1-c_W^2$. Following the standard notation, the $\gamma-\tilde{\chi}^+_i-
\tilde{\chi}^-_j$, $W^\pm-\tilde{\chi}^0_i-\tilde{\chi}^\pm_j$,
$Z-\tilde{\chi}^+_i-\tilde{\chi}^-_j$, and $Z-\tilde{\chi}^0_i-
\tilde{\chi}^0_j$ interaction vertices shown in Fig.\ \ref{fig:b1} are
%
%
proportional to $\delta_{ij}$ and \cite{Haber:1984rc}
\bea
 O^L_{ij} =
 -\frac{c_W}{\sqrt{2}} N_{i4} V^\ast_{j2} + c_W N_{i2} V^\ast_{j1} &{\rm
 ~~~~and~~~~}& O^R_{ij} = \frac{c_W}{\sqrt{2}} N_{i3}^\ast U_{j2} +
 c_W N_{i2}^\ast U_{j1},~\nonumber \\
 O^{\prime L}_{ij} = -V_{i1} V_{j1}^\ast -
 \frac{1}{2} V_{i2} V_{j2}^\ast + \delta_{ij} x_W &{\rm
 ~~~~and~~~~}& O^{\prime R}_{ij} = -U_{i1}^\ast U_{j1} -
 \frac{1}{2} U_{i2}^\ast U_{j2} + \delta_{ij} x_W,~\nonumber \\
 O^{\prime\prime L}_{ij} = -\frac{1}{2} N_{i3} N_{j3}^\ast +
 \frac{1}{2} N_{i4}N_{j4}^\ast &{\rm ~~~~and~~~~}& O^{\prime\prime
 R}_{ij} = \frac{1}{2}
 N_{i3}^\ast N_{j3} - \frac{1}{2} N_{i4}^\ast N_{j4}.
\eea

\begin{figure}
 \centering
 \epsfig{file=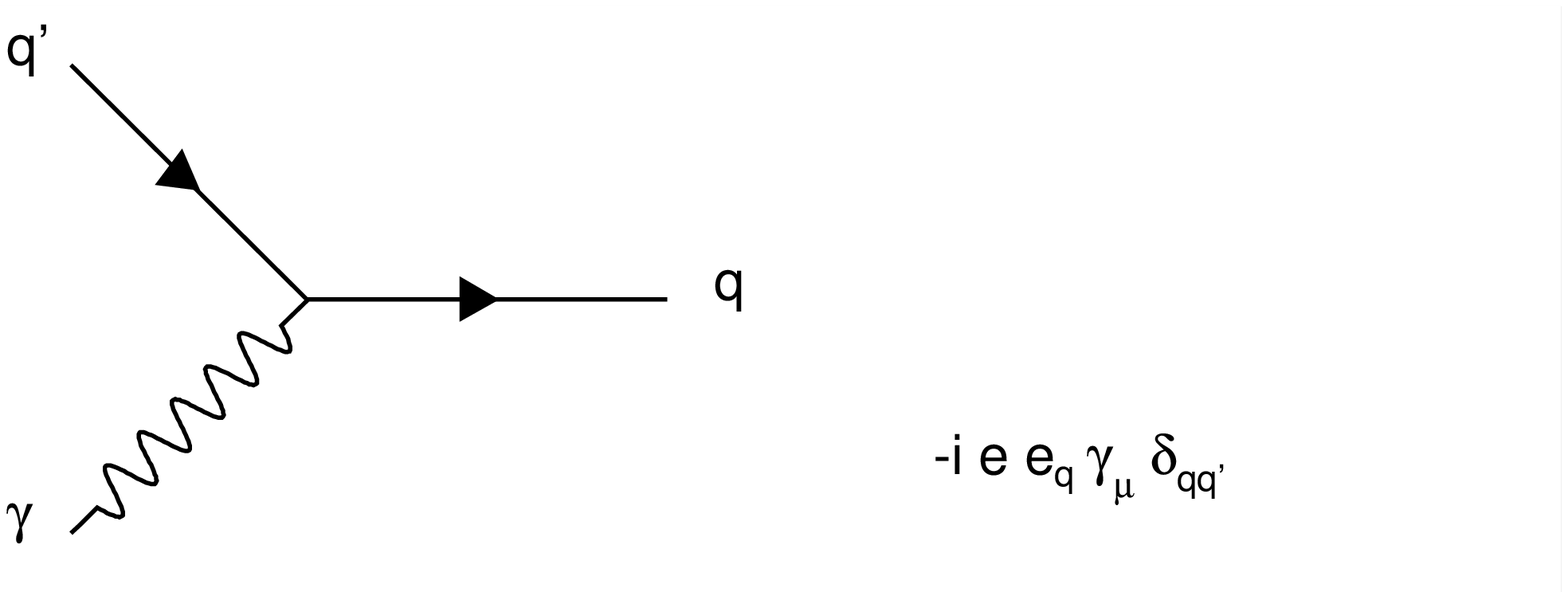,width=.49\columnwidth}
 \epsfig{file=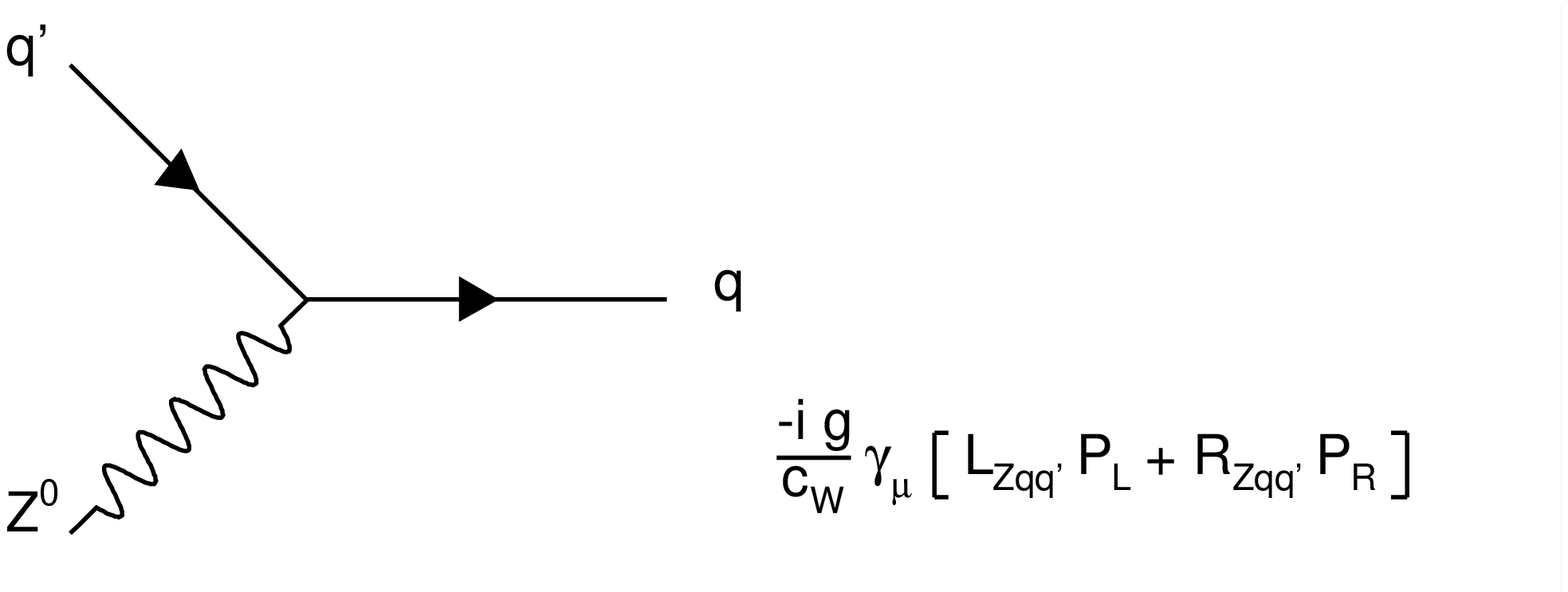,width=.49\columnwidth}\\
 \epsfig{file=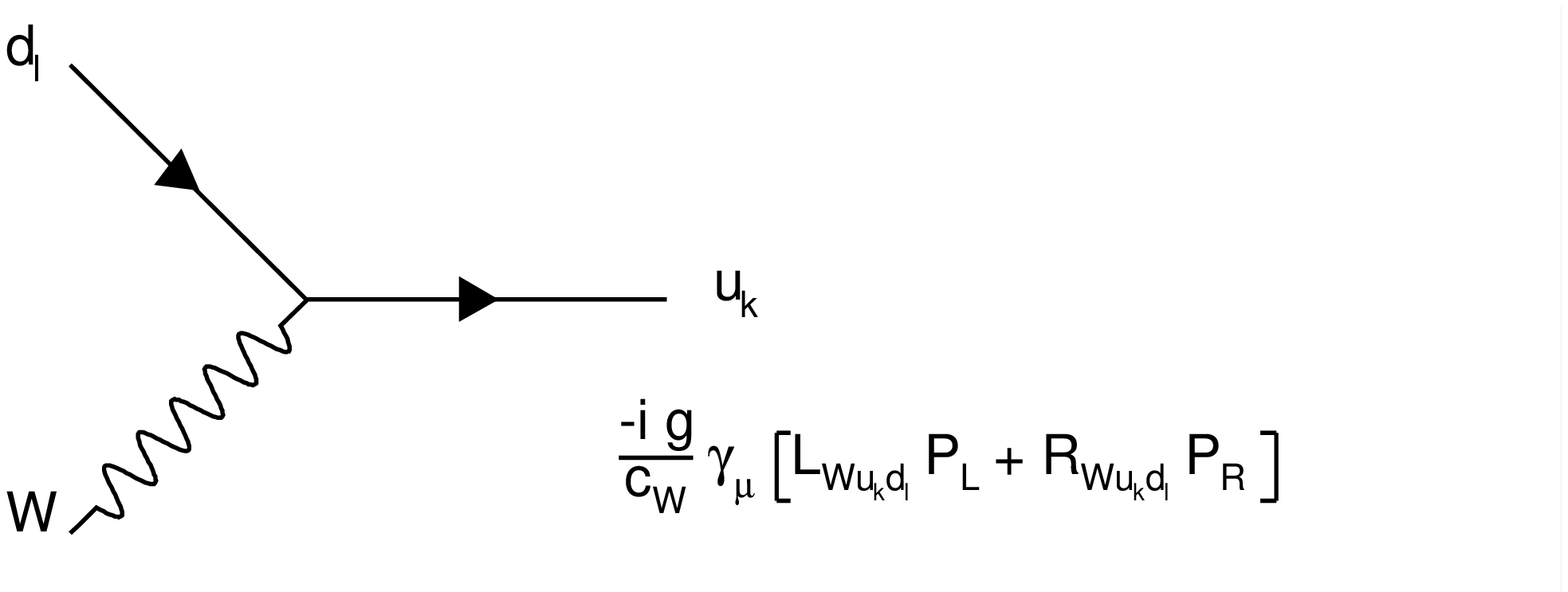,width=.49\columnwidth}
 \caption{\label{fig:b2}Feynman rules for interactions of left- and
 right-handed quarks with neutral (top) and charged (bottom) electroweak
 gauge bosons.}
\end{figure}

The interaction vertices of left- and right-handed quarks with electroweak
gauge bosons shown in Fig.\ \ref{fig:b2}
%
%
are proportional to
\bea
 \{ L_{Zq q^\prime},R_{Zqq^\prime} \}&=& (T^{3}_q -
 e_q\,x_W) \times \delta_{q q^\prime},~ \nonumber\\
 \{L_{Wqq^{\prime}},R_{Wqq^{\prime}}\} &=&
 \{c_W\,V_{qq^{\prime}}/\sqrt{2}, 0\},~
\eea
where the weak isospin quantum numbers are $T_q^3 = \pm1/2$ for left-handed
and $T_q^3=0$ for right-handed up- and down-type quarks, their fractional
electromagnetic charges are denoted by $e_q$, and $V_{qq'}$ are the elements
of the CKM-matrix. To simplify the notation, we have introduced flavor
indices in the latter, $d_1=d$, $d_2=s$, $d_3=b$, $u_1=u$, $u_2=c$, and
$u_3=t$.

\begin{figure}
 \centering
 \epsfig{file=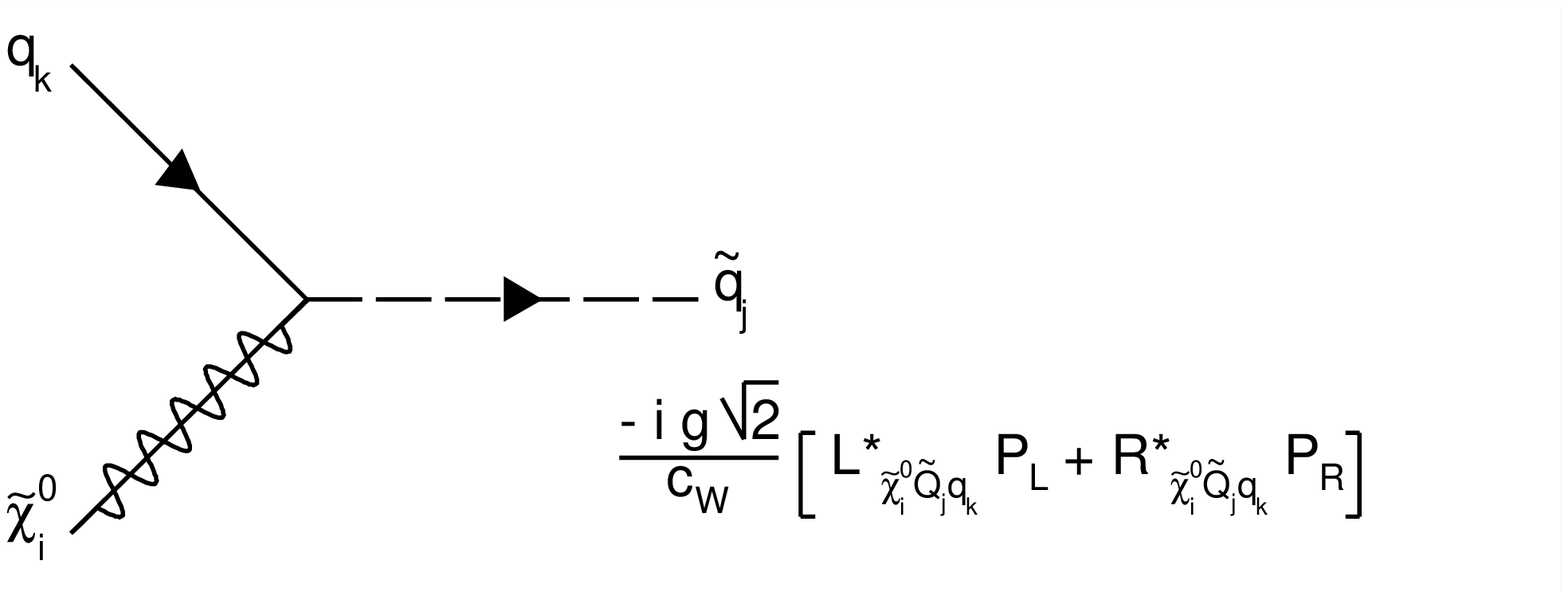,width=.49\columnwidth}\\
 \epsfig{file=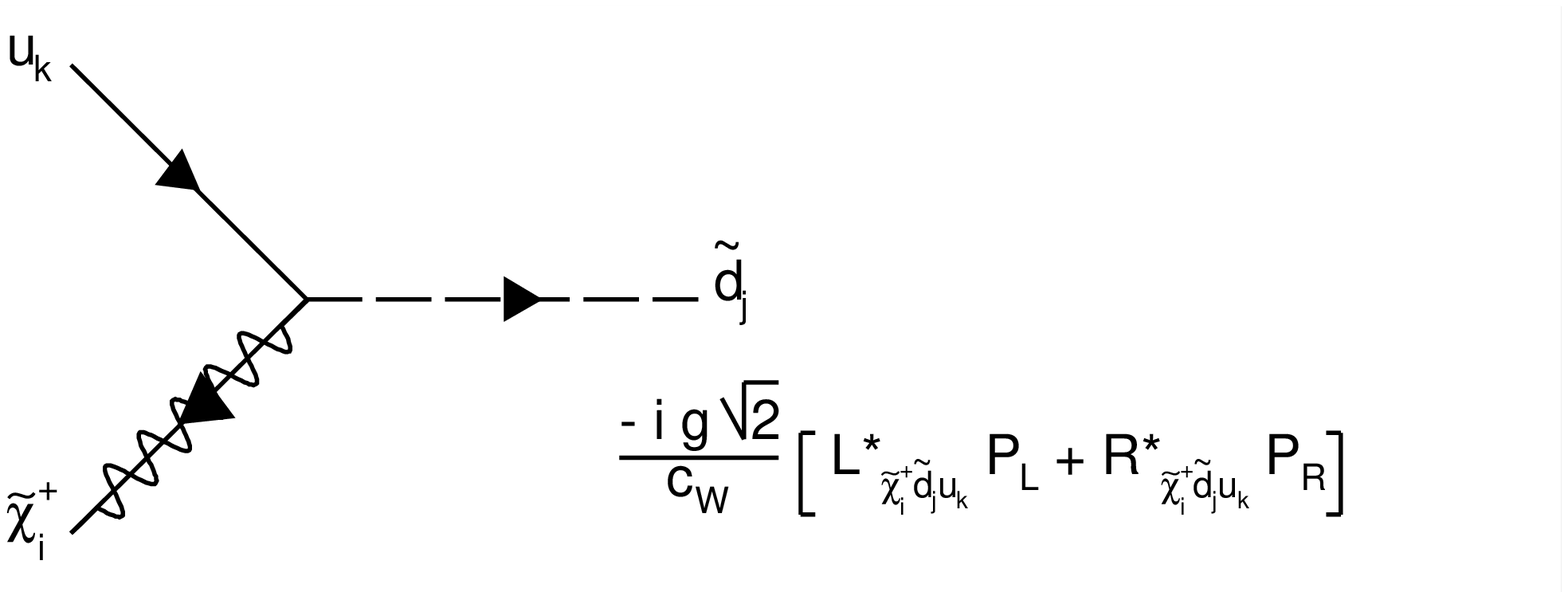,width=.49\columnwidth}
 \epsfig{file=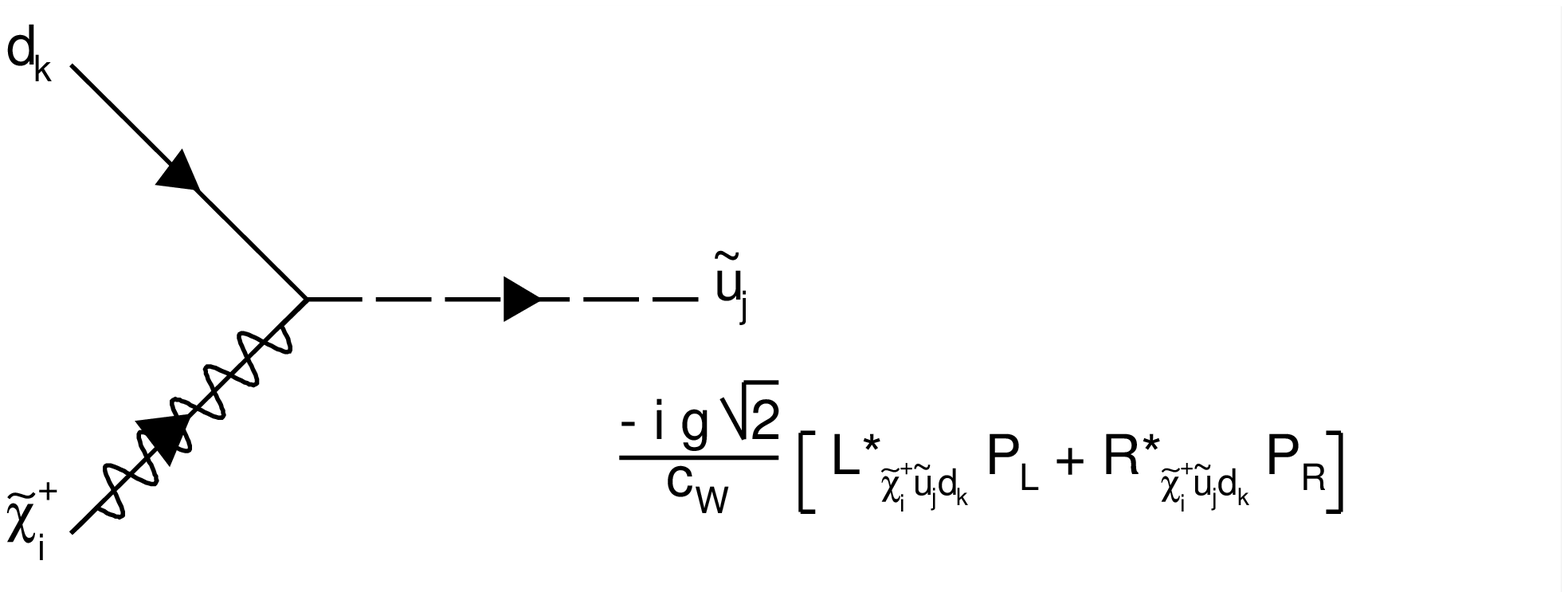,width=.49\columnwidth}
 \caption{\label{fig:b3}Feynman rules for interactions of left- and
 right-handed (s)quarks with neutral (top) and charged (bottom) gauginos.}
\end{figure}

The SUSY counterparts of these vertices correspond to the
quark-squark-gaugino vertices shown in Fig.\ \ref{fig:b3}.
%
%
These couplings are proportional to \cite{Gunion:1984yn,Rosiek:1989rs}
\bea
 L_{\tilde{\chi}^0_i\tilde{d}_j d_k} &=& \bigg[ (e_q - T^3_q)\, s_W\, N_{i1}
 +T^3_q\, c_W\, N_{i2} \bigg] R^{d\ast}_{jk}+ \frac{m_{d_k}\, c_W\,N_{i3}\,
 R^{d\ast}_{j(k+3)}}{2\, m_W\, c_\beta},~  \nonumber\\
 -R_{\tilde{\chi}_i^0\tilde{d}_j d_k }^\ast &=& e_q\, s_W\,
 N_{i1}\, R^d_{j(k+3)} - \frac{m_{d_k}\, c_W\, N_{i3}\,
 R^d_{jk}}{2\, m_W\, c_\beta} ,~\nonumber\\
 L_{\tilde{\chi}^0_i\tilde{u}_j u_k} &=& \bigg[ (e_q - T^3_q)\, s_W\, N_{i1}
 +T^3_q\, c_W\, N_{i2} \bigg] R^{u\ast}_{jk}+ \frac{m_{u_k}\, c_W\,
 N_{i4}\, R^{u\ast}_{j(k+3)}}{2\, m_W\, s_\beta}  ,~\nonumber\\
 -R_{\tilde{\chi}_i^0\tilde{u}_j u_k }^\ast &=& e_q\, s_W\,
 N_{i1}\, R^u_{j(k+3)} - \frac{m_{u_k}\, c_W\, N_{i4}\,
 R^u_{jk}}{2\, m_W\, s_\beta} ,~\nonumber \\
 L_{\tilde{\chi}_i^+\tilde{d}_j u_l }&=&
 \sum_{k=1}^3\bigg[ {c_W\over\sqrt{2}} \,U_{i1}\, R^{d\ast}_{jk}
 - \frac{m_{d_k}\,c_W\, U_{i2}\, R^{d\ast}_{j(k+3)}}{2\, m_W\,
 c_\beta} \bigg] V_{u_l d_k} ,~\nonumber\\
 -R^\ast_{\tilde{\chi}_i^+\tilde{d}_j u_l } &=&
 \sum_{k=1}^3 \frac{m_{u_l}\,c_W\, V_{i2}\,
 V_{u_l d_k}^\ast\, R^d_{jk}}{2\, m_W\, s_\beta} ,~\nonumber\\
 L_{\tilde{\chi}_i^+\tilde{u}_j d_l }&=& \sum_{k=1}^3 \bigg[
 {c_W\over\sqrt{2}}\,V_{i1}\, R^{u\ast}_{jk} - \frac{m_{u_k}\,c_W\, V_{i2}\,
 R^{u\ast}_{j(k+3)}}{2\, m_W\, s_\beta}  \bigg] V_{u_k d_l}^\ast
 ,~\nonumber \\
 -R^\ast_{\tilde{\chi}_i^+\tilde{u}_j d_l } &=&
 \sum_{k=1}^3 \frac{m_{d_l}\,c_W\, U_{i2} V_{u_k d_l}\,
 R^{u}_{jk}}{2\, m_W\, c_\beta}~,\label{eq:coup1}
\eea
where the matrices $N$, $U$ and $V$ relate to the gaugino/Higgsino mixing
(see App.\ \ref{sec:a}). All other couplings vanish due to (electromagnetic)
charge conservation (e.g.\ $L_{\tilde{\chi}_i^+\tilde{u}_j u_l }$). These
general expressions can be simplified by neglecting the Yukawa couplings
except for the one of the top quark, whose mass is not small compared to
$m_W$. Fermion number violating interactions are treated using the rules in
Ref.\ \cite{Denner:1992vz}.

In general SUSY models with non-minimal flavor violation
\cite{Gabbiani:1988rb}, the diagonalization of the mass squark matrices
$M_{\tilde{u}}^2$ and $M_{\tilde{d}}^2$ requires the introduction of two
additional $6 \times 6$ matrices $R^u$ and $R^d$ with
\bea
 {\rm diag}\,(m_{\tilde u_1}^2, \ldots, m_{\tilde u_6}^2) ~=~ R^u\,
 M_{\tilde{u}}^2\,R^{u\dag} &{\rm ~~~~and~~~~}&
 {\rm diag}\,(m_{\tilde d_1}^2, \ldots, m_{\tilde d_6}^2) ~=~ R^d\,
 M_{\tilde{d}}^2\, R^{d\dag}.
\eea
By convention, the masses are ordered according to $m_{\tilde q_1} < \ldots
< m_{\tilde q_6}$. The physical mass eigenstates are given by
\bea
 \begin{pmatrix} \tilde{u}_1 \\ \tilde{u}_2 \\
 \tilde{u}_3 \\ \tilde{u}_4 \\ \tilde{u}_5 \\ \tilde{u}_6 \\
 \end{pmatrix} = R^{u} \begin{pmatrix} \tilde{u}_L \\ \tilde{c}_L
 \\ \tilde{t}_L \\ \tilde{u}_R \\ \tilde{c}_R \\ \tilde{t}_R \\
 \end{pmatrix} &{\rm ~~~~and~~~~} \begin{pmatrix} \tilde{d}_1 \\
 \tilde{d}_2 \\ \tilde{d}_3 \\ \tilde{d}_4 \\ \tilde{d}_5 \\
 \tilde{d}_6 \\ \end{pmatrix} = R^{d} \begin{pmatrix} \tilde{d}_L
 \\ \tilde{s}_L \\  \tilde{b}_L \\ \tilde{d}_R \\ \tilde{s}_R \\
 \tilde{b}_R \\ \end{pmatrix}.
\eea
In the limit of vanishing off-diagonal parameters in the mass matrices, the
matrices $R^q$ become flavor-diagonal, leaving only the well-known
helicity mixing already present in constrained minimal flavor violation. Since we
consider numerically only the exchange of nonmixing squarks, we take
there $R^q=1$.


\end{document}